%% file: DIR_FVspectralCT.tex
\documentclass[journal,twoside,web]{ieeecolor}
\usepackage{tmi}
\usepackage{cite}
\usepackage{amsmath,amssymb,amsfonts}
\usepackage{algorithmic}
\usepackage{graphicx}
\usepackage{textcomp}

\usepackage{float,url,bm}
\usepackage{booktabs,multicol,multirow}

\DeclareMathOperator*{\argmin}{arg\,min}

\def\Real{\mathbb{R}}

\def\bx{\bm{x}}
\def\by{\bm{y}}
\def\bz{\bm{z}}
\def\bv{\bm{v}}

\newcommand{\abs}[1]{\left| #1\right|}
\newcommand{\norm}[1]{\left\lVert#1\right\rVert}

\def\BibTeX{{\rm B\kern-.05em{\sc i\kern-.025em b}\kern-.08em
    T\kern-.1667em\lower.7ex\hbox{E}\kern-.125emX}}
\begin{document}
\title{Deep Few-view High-resolution Photon-counting CT at Halved Dose for {Extremity Imaging}}
\author{Mengzhou Li, \IEEEmembership{Member, IEEE}, Chuang Niu, \IEEEmembership{Member, IEEE}, Ge Wang, \IEEEmembership{Fellow, IEEE}, Maya R Amma, Krishna M Chapagain, Stefan Gabrielson, Andrew Li, Kevin Jonker, Niels de Ruiter, Jennifer A Clark, Phil Butler, Anthony Butler, and Hengyong Yu, \IEEEmembership{Fellow, IEEE}
\vspace{-10truemm}
\thanks{This work was supported in part by NIH under the grants R01CA237267, R01EB034737, R01CA233888, R01HL151561, and R01EB031102. %
M. Li, C. Niu, and G. Wang are with the Biomedical Imaging Center, Rensselaer Polytechnic Institute, Troy, New York, USA (e-mail:mzli@ieee.org; niuc@rpi.edu; wangg6@rpi.edu). %
M. R. Amma, N. d. Ruiter, and P. Butler are affiliated with MARS Bioimaging Limited, Christchurch, New Zealand (e-mail: maya.vivek1@gmail.com; niels.deruiter@marsbioimaging.com; phil.butler@marsbioimaging.com).  %
K. M. Chapagain is with MARS Bioimaging Limited, Christchurch, New Zealand, also with University of Otago, New Zealand (e-mail: krish22chapagain@gmail.com). %
S. Gabrielson is with Canterbury District Health Board, New Zealand (e-mail: STEFAN.GABRIELSON@cdhb.health.nz). %
A. Li is with Pacific Radiology, New Zealand (e-mail: andrew.li@pacificradiology.com).  %
K. Jonker is with MARS Bioimaging Limited, Christchurch, New Zealand, also with University of Canterbury, New Zealand (e-mail: kevin.jonker@marsbioimaging.com). %
J. A. Clark is with MARS Bioimaging Limited, Christchurch, New Zealand, also with University of Otago, New Zealand (e-mail: jenn.clark@marsbioimaging.com). %
A. Butler is with MARS Bioimaging Limited, Christchurch, New Zealand, also with University of Otago, New Zealand, University of Canterbury, New Zealand, and Canterbury District Health Board, New Zealand (e-mail: anthony.butler@marsbioimaging.com). %
H. Yu is with Department of Electrical and Computer Engineering, University of Massachusetts Lowell, Lowell, Massachusetts, USA (e-mail: Hengyong-yu@ieee.org). The corresponding authors are G. Wang, A. Butler, and H. Yu. }}

\maketitle
\begin{abstract}
X-ray photon-counting computed tomography (PCCT) for extremity allows multi-energy high-resolution (HR) imaging but its radiation dose can be further improved.
Despite the great potential of deep learning techniques, their application in HR volumetric PCCT reconstruction has been challenged by the large memory burden, training data scarcity, and domain gap issues.
In this paper, we propose a deep learning-based approach for PCCT image reconstruction at halved dose and doubled speed validated in a New Zealand clinical trial. Specifically, we design a patch-based volumetric refinement network to alleviate the GPU memory limitation, train network with synthetic data, and use model-based iterative refinement to bridge the gap between synthetic and clinical data. Our results in a reader study of 8 patients from the clinical trial demonstrate a great potential to cut the radiation dose to half that of the clinical PCCT standard without compromising image quality and diagnostic value.
\end{abstract}

\begin{IEEEkeywords}
Photon-counting CT, few-view reconstruction, radiation dose reduction, deep learning, clinical trial.
\vspace{-6truemm}
\end{IEEEkeywords}

\section{Introduction}
\label{sec:introduction}
\IEEEPARstart{G}iven the potential patient risk under ionizing radiation, research is actively performed to reduce computed tomography (CT) radiation dose\cite{joyce2020strategies}. For example, we can optimize scanning parameters for different patients and use automatic exposure control~\cite{soderberg2010automatic}. Recent development of photon-counting CT (PCCT) and deep learning-based reconstruction algorithms gives new promises in this regard~\cite{muhammad2022impact, willemink2018photon}. 

The great potential of PCCT in clinical utilities has been well demonstrated in atherosclerosis imaging, extremity scanning, and multi-contrast-enhanced studies \cite{si2022coronary,mergen2022first,panta2018first,stamp2019clinical,ostadhossein2020multi}.  
Since 2019, the PCCT company MARS has been conducting human clinical trials for orthopaedic and cardiovascular applications, and already expanded the trials into the local acute care clinics. The orthopaedic trials have shown that PCCT imaging at a high resolution is advantageous in the acute, follow-up, pre-surgical and post-surgical stages. Efforts are being made to conduct clinical trials in Europe for rheumatology applications.

Despite the huge potential of extremity PCCT, a few challenges must be addressed for further improvements~\cite{wang2020spectral}. The current low-cost and light-weight specialty MARS CT scanner uses neither an expensive gantry nor a heavy X-ray source. A customized gantry offers cost-effectiveness at a slow scanning speed. A micro-focus X-ray tube improves spatial resolution but, at the same time, limits the photon flux. The mechanical limitations and low photon flux introduce significant challenges in temporal resolution and noise. First, the MARS PCCT scanner currently takes over 8 minutes to scan a patient, which cannot support dynamic contrast-enhanced studies. Second, channel-wise projections suffer from low signal-to-noise ratios. For instance, with our current protocol, fewer than 1,500 photons are split into five non-overlapping energy bins, resulting in only hundreds of photons in one channel as opposed to \(\sim 1\times10^5\) photons for conventional CT. This becomes more problematic with a narrow energy bin. To mitigate these issues, a natural solution is to reduce the number of projection views per scan and to develop advanced reconstruction techniques.

Few-view and low-dose CT reconstruction is a main area of CT research. In the early stages, compressed sensing was widely used, such as total variation (TV) and dictionary learning techniques~\cite{sidky2006accurate,makinen2020collaborative}. More recently, deep learning methods 
delivered exciting reconstructions~\cite{Wang2020NMA}, becoming the new frontier along the direction. However, there are still several gaps to meet for high-resolution (HR) PCCT. First, the existing methods are mainly developed for CT image reconstruction in single-channel mode and 2D imaging geometry, few of which target on volumetric reconstruction in multi-channel mode at high resolution due to GPU memory constraints~\cite{he2018optimizing,chen2018learn,shen2019patient,wu2021drone,thies2022learned,li2023noise}. Second, it is well known that the network performance could drop significantly if the data used during inference comes from a distribution different from that of the training data~\cite{li2023realistic,du2021investigation,xu2020noisy}. This domain gap is more critical for diagnostic image reconstruction as it is less tolerant of artifacts and ``hallucinations'' (generating factually incorrect or misleading structures with realistic appearances) than in some other fields. Third, high quality datasets are scarce for network training since HR PCCT is an emerging technology. Although some unsupervised and self-supervised methods work without paired data, they often rely on specific assumptions about noise characteristics in the images, do not consider the inter-channel correlation in spectral images, and demonstrate sub-optimal performance~\cite{moran2020noisier2noise,hendriksen2020noise2inverse,xu2020noisy,yuan2020half2half,zhang2021noise2context,niu2022noise}. 

In this paper, we present a deep learning-based approach to address the aforementioned challenges in a few-view mode at a halved dose (using half the number of projections from the original full-dose scan, with ``half-view" and ``half-dose" used interchangeably hereafter unless explicitly noted) relative to the commercial PCCT technology used in the New Zealand clinical trial. 
We summarize our primary contributions as follows:
\begin{itemize}
  \item We develop a deep learning-based reconstruction pipeline for volumetric HR PCCT reconstruction. The pipeline is memory-efficient on a single workstation, aided by a shared low noise prior for all channels, patch-based deep iterative refinement on channel-wise volume reconstruction, and texture tuning that leverages inter-channel correlations for full-channel slices;
  \item We demonstrate the potential in addressing the domain gap issues using our patch-based volumetric denoising combined with a model-based iterative refinement framework. With the proposed network trained on synthetic data, we consistently achieve excellent results on both phantom data and patient scans acquired on different machines with different protocols;
  \item Our half-view/half-dose PCCT reconstruction results are favored by radiologists over the proprietary reconstruction from the full-view dataset in terms of diagnostic image quality, suggesting a great potential for clinical translation to address data scarcity.
\end{itemize}
To the best of our knowledge, this is the first attempt at deep learning-based volumetric reconstruction for multi-channel HR PCCT, \emph{e.g.}, $1,200^3\times5$. This also represents the first study reporting superior diagnostic quality at half-dose with a deep network trained on synthetic data over full-dose clinical PCCT reconstruction.

\section{Methods}

\begin{figure*}
\vspace{-3truemm}
  \centering
  \includegraphics[width=0.8\linewidth]{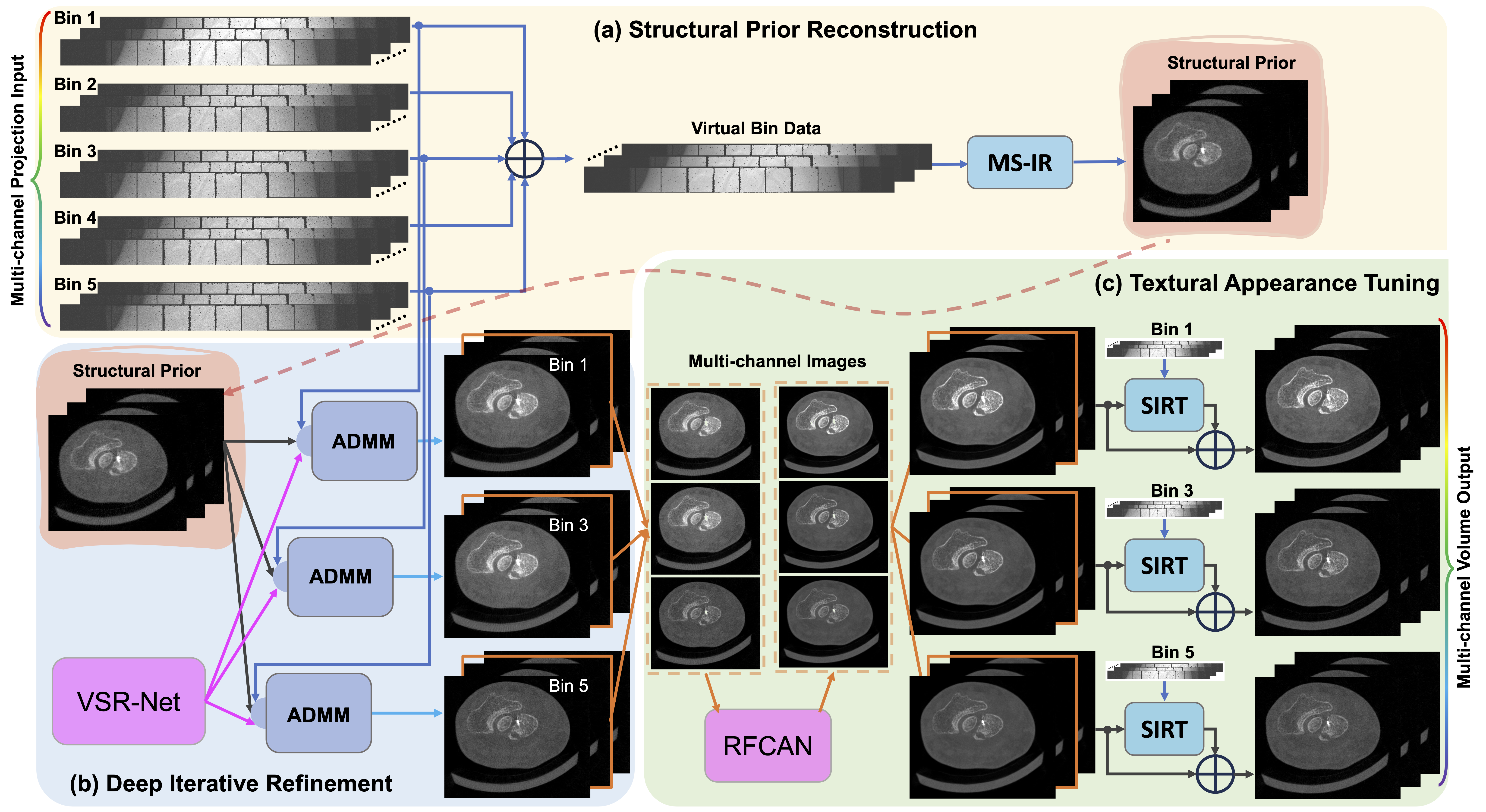}
  \vspace{-3truemm}
  \caption{{Deep few-view PCCT workflow.} (a) A less noisy structural prior is reconstructed by summing counts from all channels and using a multi-scale iterative reconstruction (MS-IR) technique; (b) for image reconstruction in each channel, the structural prior is iteratively refined using a Volumetric Sparse Representation Network (VSR-Net) and model-based guidance with the projection measurements in an Alternating Direction Method of Multipliers (ADMM) framework; and (c) the multi-channel images are further refined using a Residual Fourier Channel Attention Network (RFCAN) for alignment with the MARS full-dose reconstruction, and followed by further polishing with the Simultaneous Iterative Reconstruction Technique (SIRT) to generate similar image sharpness and noise characteristics that radiologists prefer.}
  \label{fig:Diagram}
  \vspace{-6truemm}
\end{figure*}

An overview of our approach is shown in Fig.~\ref{fig:Diagram}, mainly consisting of three parts: structural prior reconstruction, deep iterative refinement, and textural appearance tuning. 
The details are elaborated in the following subsections.

\subsection{MARS Extremity PCCT}
The clinical trial was performed on the state-of-the-art MARS Extremity 5X120 scanner, which can simultaneously measure five effective energy bins at spatial resolution \(50-200 \mu m\). The system includes CdTe-Medipix3RX photon-counting detectors (PCDs) with \(110 \mu m\) pixel pitch (12 chips arranged in a non-flat arc shape), an X-ray source (up to \(120 kVp\), \(350 \mu A\)), and a rotating gantry for helical scanning. It provides isotropic \(90^3 \mu m^3\) voxel size. The bore size is \(125 mm\) with a scanning length of \(35cm\) for extremity scans.

\subsection{Reconstruction with Structural Prior}\label{sec:StructurePrior}
Each element of the MARS PCD counts with 5 effective energy thresholds simultaneously, resulting in quasi-monochromatic projections in 5 energy bins: \emph{i.e.}, \(7-40 keV\), \(40-50 keV\), \(50-60 keV\), \(60-70 keV\), and \(70 keV\) above.
The patient is scanned with a total incoming photon count of around 1,500 per detector element for open-beam measurement, resulting in only hundreds of photons in one channel.
Given such low counts, direct reconstruction in each energy bin inevitably suffers from major quantum noise. Instead, we employ a similar idea from the prior image constrained compressed sensing~\cite{chen2008prior} and its extension to PCCT~\cite{rajendran2018ultra}, by noticing that the structural information among different energy bins is closely correlated, with only slight differences in attenuation values. Hence, we propose the following steps for spectral reconstructions in 5 energy bins with minimized quantum noise:
  (1) We sum the counts from all channels to form a virtual `integrating' bin with minimized quantum uncertainty;
  (2) We reconstruct from the virtual bin to obtain an image as a low-noise structural prior;
  (3) Leveraging the inter-bin similarity, we initialize our iterative deep reconstruction method with the structural prior, and feed in bin-wise real data to reconstruct the spectral image. A multi-scale iterative reconstruction strategy is used to significantly accelerate the convergence for the large volume reconstruction.
Note that we enforce this structural similarity here by using the prior only as an initialization for the best computational efficiency in contrast to traditional approaches~\cite{rajendran2018ultra},  which use the prior as a constraint and solve the optimization problem iteratively. As a cost, our approach in this step does not guarantee a solid convergence. Hence, we rely on iterative reconstruction constrained by a deep neural network-based prior regularization for valid image reconstruction as introduced in the following section, while this virtual-bin structural prior initialization serves as an acceleration step.

\subsection{Deep Iterative Refinement (DIR)}
To address the challenge of data scarcity, we propose to use synthetic data for network training and address the domain gap issue using the following strategies. First, we limit the function of network to low-level feature denoising, which is less sensitive to domain gaps. Second, a patch-based training strategy is employed to leverage low-level similarity and reduce the domain gap. Furthermore, model-based iterative refinement is used to suppress remaining errors. These elements are integrated using the alternating direction method of multipliers (ADMM). 

\subsubsection{ADMM Optimization}
The solution space under a data constraint is often high-dimensional for a few-view or low-dose CT reconstruction problem. Mathematically, this formulates an optimization problem as follows:
\begin{equation}\label{eq:opt_problem}
  \bx^* = \argmin_{\bx}\frac{1}{2}{\norm{A\bx - \by}^2 + \lambda R(\bx)}, %
\end{equation}
where $A\in \Real^{M\times N}$ and $\by \in \Real^{M}$ are a system matrix and a measurement vector respectively, $\bx \in \Real^N$ denotes an image volume to be reconstructed, and $R(\cdot)$ is a regularization term. 

To solve Eq.~(\ref{eq:opt_problem}) with deep prior, an auxiliary variable $\bz$ is introduced to decouple the prior term from the loss function as follows:
\begin{equation}
  \bx^* = \argmin_{\bx}\frac{1}{2}{\norm{A\bx - \by}^2 + \lambda R(\bz)}, \quad \text{s.t.} \quad \bz = \bx. \label{eq:optconstrained}
\end{equation}
The augmented Lagrangian~\cite{hestenes1969multiplier} of Eq.~(\ref{eq:optconstrained}) is written as
\begin{equation}
  \mathcal{L}_\mu(\bx,\bz,\bv) = \frac{1}{2}\norm{A\bx - \by}^2 + \lambda R(\bz) + \bm{v}^T(\bx-\bz) +\frac{\mu}{2}\norm{\bx-\bz}^2, \label{eq:ADMM_Lag}
\end{equation}
which becomes a saddle point problem and can be solved using the ADMM~\cite{boyd2011distributed,wang2019global} as follows:

\begin{equation}
  \left\{\begin{array}{l}
    \bx^{k+1} = \argmin_{\bx} \frac{1}{2}\norm{A\bx - \by}^2 +\frac{\mu}{2}\norm{\bx-\bz^k + \frac{\bv^k}{\mu}}^2 \\
    \bz^{k+1} = \argmin_{\bz} \lambda R(\bz) +\frac{\mu}{2}\norm{\bx^{k+1}-\bz + \frac{\bv^k}{\mu}}^2               \\
    \bv^{k+1} = \bv^{k} + \mu (\bx^{k+1} - \bz^{k+1})                                                            \\
  \end{array} ,\label{eqn:ADMM}
  \right.
\end{equation}
where $\mu$ is a hyper-parameter and $\bv$ is the augmented Lagrange multiplier.

The optimization of $\bx$ can be achieved using the gradient descent method for a number of steps with a step size $\beta$:
\begin{equation}
  \begin{aligned}
    \nabla \mathcal{L}_\mu(\bx) & = A^T(A\bx - \by) + \mu(\bx - \bz^k + \bv^k/\mu),           \\
    \bx^{k, (t+1)}              & = \bx^{k, (t)} - \beta \nabla \mathcal{L}_\mu(\bx^{k, (t)}),
  \end{aligned}
\end{equation}
where $t$ represents the step index for the gradient descent search. The optimization of $\bz$ is a proximal operation:
\begin{equation}
  \bz^{k+1} = \text{prox}_{\frac{\lambda}{\mu}R}(\bx^{k+1} + \bv^k/\mu).\label{eq:proximalOP}
\end{equation}
Clearly, a learned denoiser resembles a projection of the noisy input onto a clean image manifold~\cite{alain2014regularized}, also shown by several recent studies using deep networks as learned proximal operators~\cite{dong2018denoising, zhang2021plug}. 
Here we use our network to approximate the proximal operation as a deep prior.

Note that the noise characteristics could change through iterations. However, the network denoiser is often trained for mapping a noisy CT reconstruction to its corresponding clean label. To reduce noise mismatch and accelerate convergence, we initialize $\bx$ with the results obtained with the structural prior. 
While the magnitude of noise in the reconstruction gradually reduces through iterations, we regulate the network contribution at later stages. Eq.~(\ref{eq:proximalOP}) is reformulated with a network denoiser as follows:
\begin{equation}
  \bz^{k+1} = \gamma  f_{\theta}(\bx^{k+1} + \bv^k/\mu) + (1-\gamma)(\bx^{k+1} + \bv^k/\mu),
\end{equation}
where $\gamma$ controls the network contribution, which can be understood as a parameter to control the amount of noise to be removed.

\subsubsection{Volumetric Sparse Representation Network (VSR-Net)}
\begin{figure}[ht]
  \centering
  \includegraphics[width=0.85\linewidth]{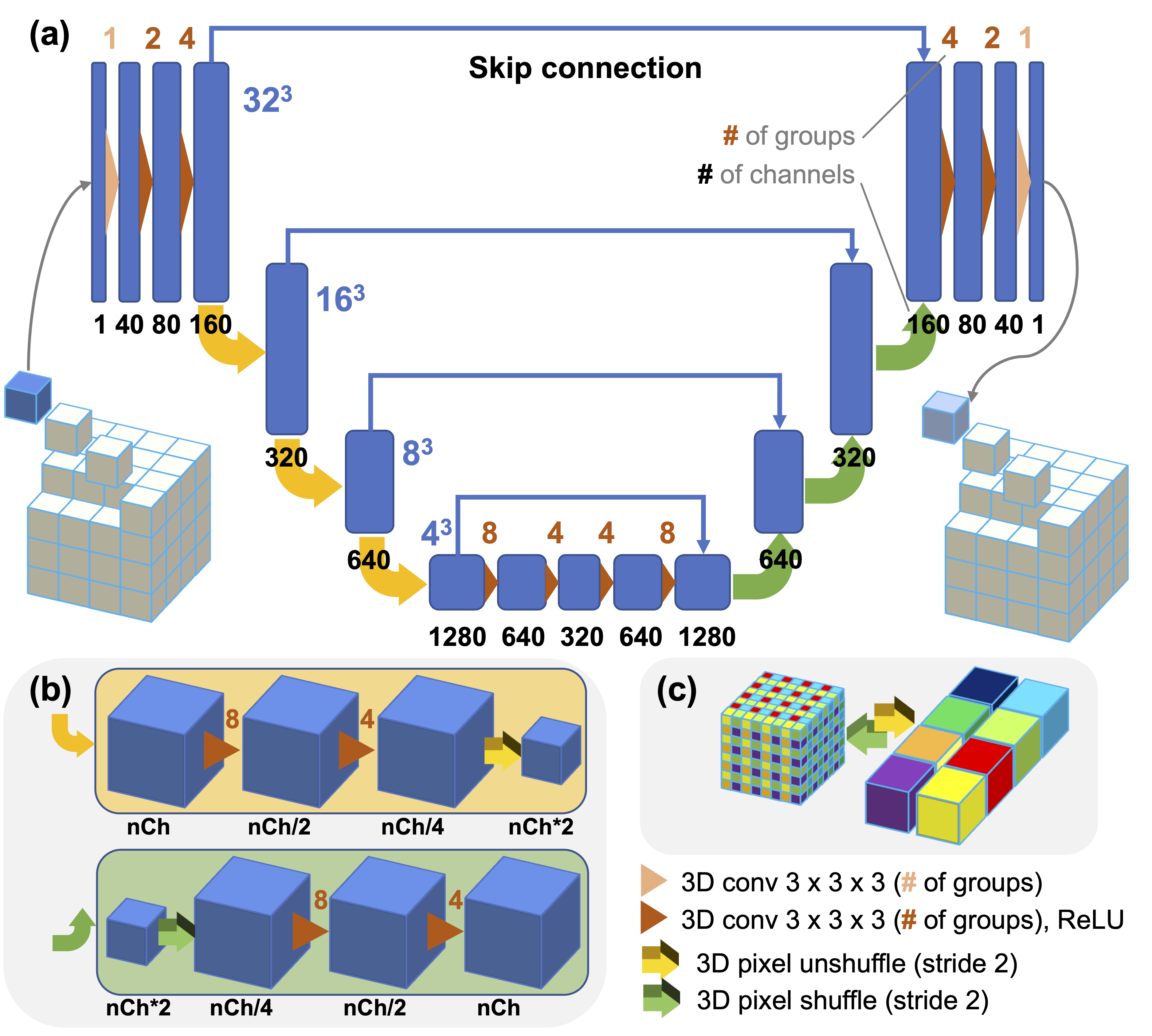}
  \vspace{-3truemm}
  \caption{{Architecture of our volumetric sparse representation network (VSR-Net).} {(a)} This light-weight network takes a small cubic patch as input and outputs a denoised patch, with 3D pixel shuffle operations and grouped convolutions; and {(b)} the downscaling and upscaling of the feature maps are achieved through 3D pixel unshuffle and shuffle operations (illustrated in (c)) combined with two 3D grouped convolutional layers. 
  Note that  a color-coded number above each convolutional operation denotes the number of groups used, while the number underneath the feature map indicates the number of channels.
  }
  \label{fig:Network}
  \vspace{-4truemm}
\end{figure}

Despite the great successes of deep 2D CT reconstruction methods, directly applying them to clinical HR PCCT for volumetric reconstruction is infeasible on conventional GPUs. The GPU memory cost becomes huge for image volume and sinogram storage as well as for the corresponding backward/forward projection operations, invalidating direct methods like AUTOMAP~\cite{zhu2018image} and other unrolling methods~\cite{he2018optimizing,xu2022convex}. Rather than training a network targeting a whole volume, we train a network that learns a patch-based representation to overcome the memory limit for volumetric reconstruction. The architecture of our proposed network is illustrated in Fig.~\ref{fig:Network}. It is a light-weight 3D network that combines U-Net~\cite{ronneberger2015u} and ResNet~\cite{xie2017aggregated} with 3D grouped convolutions ~\cite{huang2018condensenet} and specialized 3D pixel shuffle operations to promote application speed and performance. 

In contrast to many generative networks using large reception fields for realistic high-level feature synthesis, we force the network to concentrate on low-level features by choosing a small kernel size of $3\times 3 \times 3$ for all convolution layers to leverage low-level structural similarities and gain more tolerance to domain shift.
Compared to widely used 2D convolutions, we use a 3D grouped convolution for all convolution layers to fully utilize the information from neighboring voxels. To facilitate training, the grouped convolution encourages structural sparsity, promotes the differentiation between feature maps, and constructs a 3D network with fewer trainable parameters. Different from the conventional downscale/upscale convolutions with strides, we use 3D pixel unshuffle/shuffle operations as shown in Fig.~\ref{fig:Network}(b). The unshuffle operation splits the input volume into 8 sub-volumes and concatenates them in the channel dimension, while the shuffle operation assembles the sub-volumes into a super volume. 
In reference to the work in~\cite{lim2017enhanced}, we omit batch normalization throughout the network, and the first and last convolution layers involve no activation. For scaling invariance and generalizability, no bias is used in all layers as suggested in~\cite{zhang2021plug}. The cube size for network training can be adjusted according to the available GPU memory. We set the cube size to 32 in our experiments, and the network can be easily deployed on a conventional standard 1080Ti GPU with 11GB memory.

\subsubsection{VSR-Net Training with Synthetic Data}\label{sec:VSR-Net_training}
Since it is rather challenging to obtain the ground truth for HR PCCT scans of patients, we use synthesized data for network training. Specifically, we construct our training dataset from the open dataset for the Low-dose CT Grand Challenge~\cite{mccollough2017low}. 
We first resize the images to have isotropic voxels of \(1 mm\) along each dimension, and convert the voxel values in Hounsfield units to linear attenuation coefficients. Then, we treat volumes as digital phantoms of \(0.2^3 mm^3\) voxel size and generate noise-free projections in the MARS CT scanner geometry. Finally, the projections with quantum noise are simulated assuming 16,000 incident photons per detector element. The Simultaneous Iterative Reconstruction Technique (SIRT) is used to reconstruct images. 

The isotropic volumes and corresponding noisy reconstructions are the labels and noisy inputs for network training. Ten patient volumes are partitioned into cubes of size $32^3$ with a stride of 25 pixels along each direction. Then, the cubes are sieved to remove empty ones based on the standard deviation of pixel values. As a result, over 190,000 pairs of 3D patches are generated for training, and around 38,000 pairs for validation. The loss function consists of an $L_1$ norm for the value difference and a mean square error for the relative value difference:
\begin{equation}
  \sum_i \left[\norm{\by_i - f_{VSR}(\bx_i;\theta)}_1 +\beta_0 \norm{\frac{\by_i - f_{VSR}(\bx_i;\theta)}{\by_i + c}}^2_2\right],
\end{equation}
where \(\by_i\) and \(\bx_i\) are respectively the label patches and noisy inputs, \(f_{VSR}(\bx_i;\theta)\) corresponds the network output with trainable parameters \(\theta\), and \(c\) is a constant to avoid zero denominator. The $L_1$ norm, instead of the $L_2$ norm, is used in the first term to avoid blurring details, and the relative error is measured with the $L_2$ norm in the second term to preserve tiny structures based on our experiences~\cite{li2020x}. During training, we set the balancing hyperparameter $\beta_0$ to 1 and \(c\) to 0.1.

\subsubsection{Parallel Batch Processing \& Geometric Self-ensemble}
During the inference, a reconstruction volume is partitioned into overlapping patches and then fed into the VSR-Net. Geometric self-ensemble based on flipping and rotation is used to boost performance and suppress checker-board artifacts. To save computational time, we randomly apply 1 of 8 transforms to the reconstruction volume for each iteration, which is modified from the periodic geometric self-ensemble idea~\cite{zhang2021plug}. For acceleration, parallel processing techniques are used to distribute the workload across multiple GPUs.

\subsection{Textural Appearance Tuning}\label{sec:TextureTuning}
To match the image texture with that of the MARS commercial reconstruction radiologists are already familiar with, we adopt a two-step refinement process. First, we use a 2D convolutional network to exploit the inter-channel correlation for texture enhancement and value alignment. Multi-channel images extracted from the channel reconstructions of the same slice are fed to the network for the mapping, in a slice-by-slice manner for memory efficiency. Then, we process the reconstruction through a few SIRT iterations to enhance image sharpness and alter noise characteristics. 
To minimize potential perception bias, we further balance the sharpened but noisier result by mixing it with the original network processed result at a ratio preferred by radiologists. Specifically, we ran 30 additional SIRT iterations for our reconstruction, and the mixing weights were set to 0.75 and 0.25 for the SIRT reconstruction and the network result, respectively. This ratio was determined by presenting a set of images with various mixing ratios to a radiologist and asking them to choose the one that produced the best perceptual image quality.

\subsubsection{Residual Fourier Channel Attention Network (RFCAN)}\label{sec:FRCAN_training}
\begin{figure}[ht]
  \centering
  \includegraphics[width=0.95\linewidth]{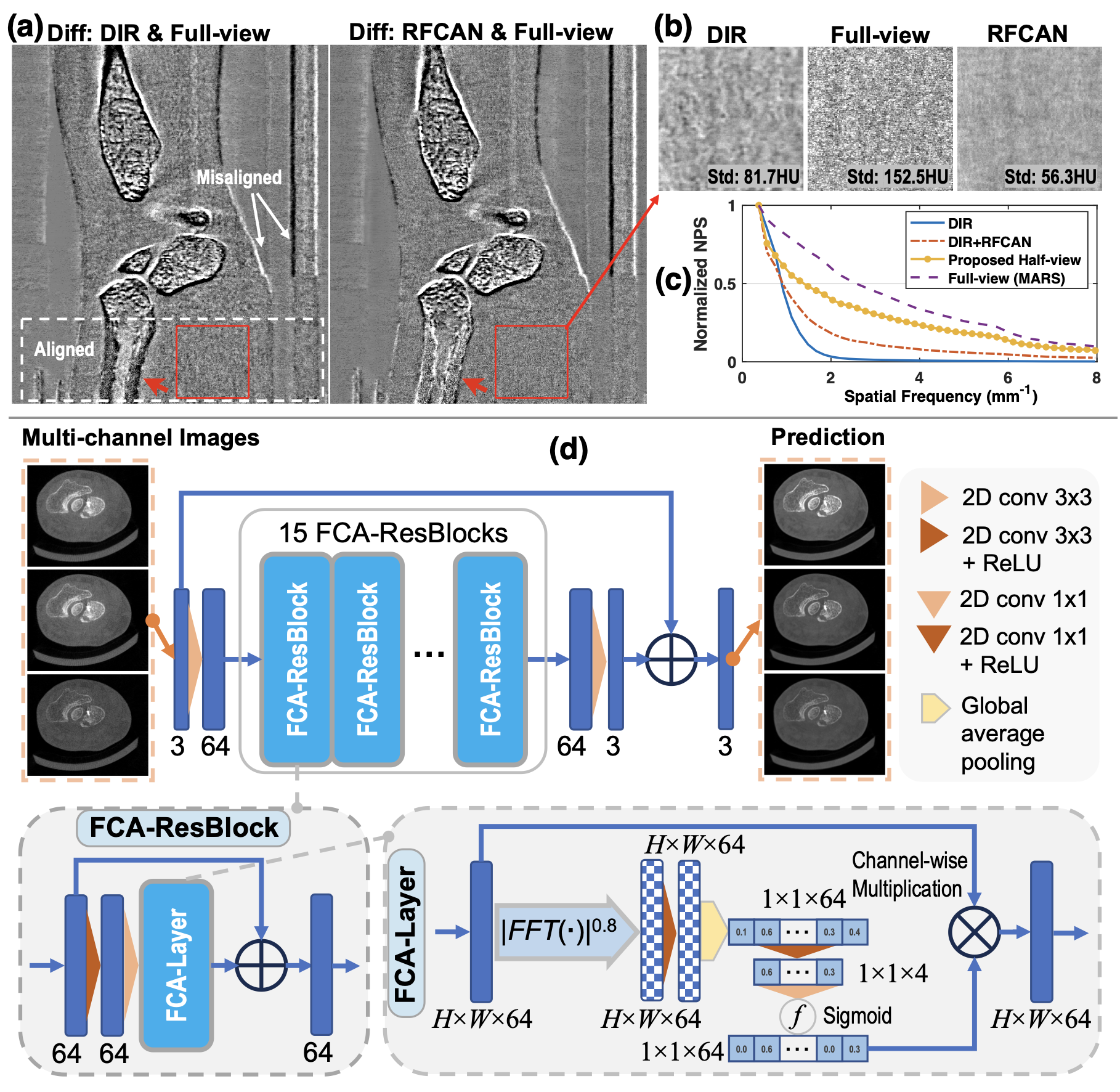}
  \vspace{-3truemm}
  \caption{Motivation and architecture of the residual Fourier channel attention network (RFCAN). It mainly intends to (i) correct contrast shift, and (ii) adjust noise texture to match clinical references. 
  (a) The difference images of deep iterative refinement (DIR) results and RFCAN results, against the MARS full-view reconstruction reference, reveal clear misregistration in the upper part of the volume---highlighting the need for a misalignment-insensitive loss function. A slight value shift for bones is also observed in the DIR result, indicated by the dark region pinpointed by arrows. (b) Zoomed-in views of a flat tissue region show a notable texture discrepancy between the DIR result and the reference, despite their similar noise levels (standard deviation) and mean values. (c) Noise power spectrum (NPS) curves, estimated from the flat tissue region, further confirm this texture difference, motivating the alignment of bone values and the adjustment of noise characteristics to reduce perception bias using RFCAN. (d) The proposed RFCAN consists of 15 Fourier channel attention residual blocks (FCA-ResBlocks) built upon the attention layers with FCA (FCA-Layer). It functions as a post-processing procedure that was applied to the multi-channel DIR outputs.
  }
  \label{fig:RFCAN}
  \vspace{-3truemm}
\end{figure}

We design a value and texture alignment network by modifying the residual channel attention network~\cite{zhang2018image} using a Fourier channel attention mechanism~\cite{qiao2021evaluation} to learn a multi-channel mapping, as illustrated in Fig.~\ref{fig:RFCAN}. 
For training the RFCAN, we use the full-view MARS reconstruction slices as the label for our half-view reconstruction results. However, due to patient motion, there can be occasional non-rigid misalignments across a few slices between our reconstruction and the MARS results, resulting from differences in volume and projection partitioning schemes, and the number of projections. Hence, we select one patient scan that is least affected by motions as our training data, and then sieve out the misalignment-affected slices, resulting in 584 pairs of HR multi-channel images ($1,200\times1,200$). A total of 206,000 pairs of overlapping patches of size $128\times128$ are randomly extracted for network training, and around 52,000 pairs for validation. An additional penalty on the Fourier spectrum (insensitive to misalignment) is introduced in the loss function to emphasize the texture similarity besides the intensity fidelity imposed by the other terms as follows: 
\begin{align}
  \sum_i \Bigl[\norm{\by_i - f(\bx_i;\theta)}_1 +\beta_1 \norm{\frac{\by_i - f(\bx_i;\theta)}{\by_i + c}}^2_2 \nonumber \\
 + \beta_2\norm{\abs{\mathcal{FFT}(\by_i)} - \abs{\mathcal{FFT}(f(\bx_i;\theta))}}_1 \Bigr],
\end{align}
where $\mathcal{FFT}(\cdot)$ denotes the Fourier transform. \(\by_i\), \(\bx_i\), and \(f(\bx_i;\theta)\) are the label, input, and network output, respectively. The balancing hyperparameters $\beta_1$ and $\beta_2$ are both set to 1 with \(c\) set to 0.1 during training.

\subsection{Interleaved Updating for Large-volume Recon}
\begin{figure}
  \centering
  \includegraphics[width=0.8\linewidth]{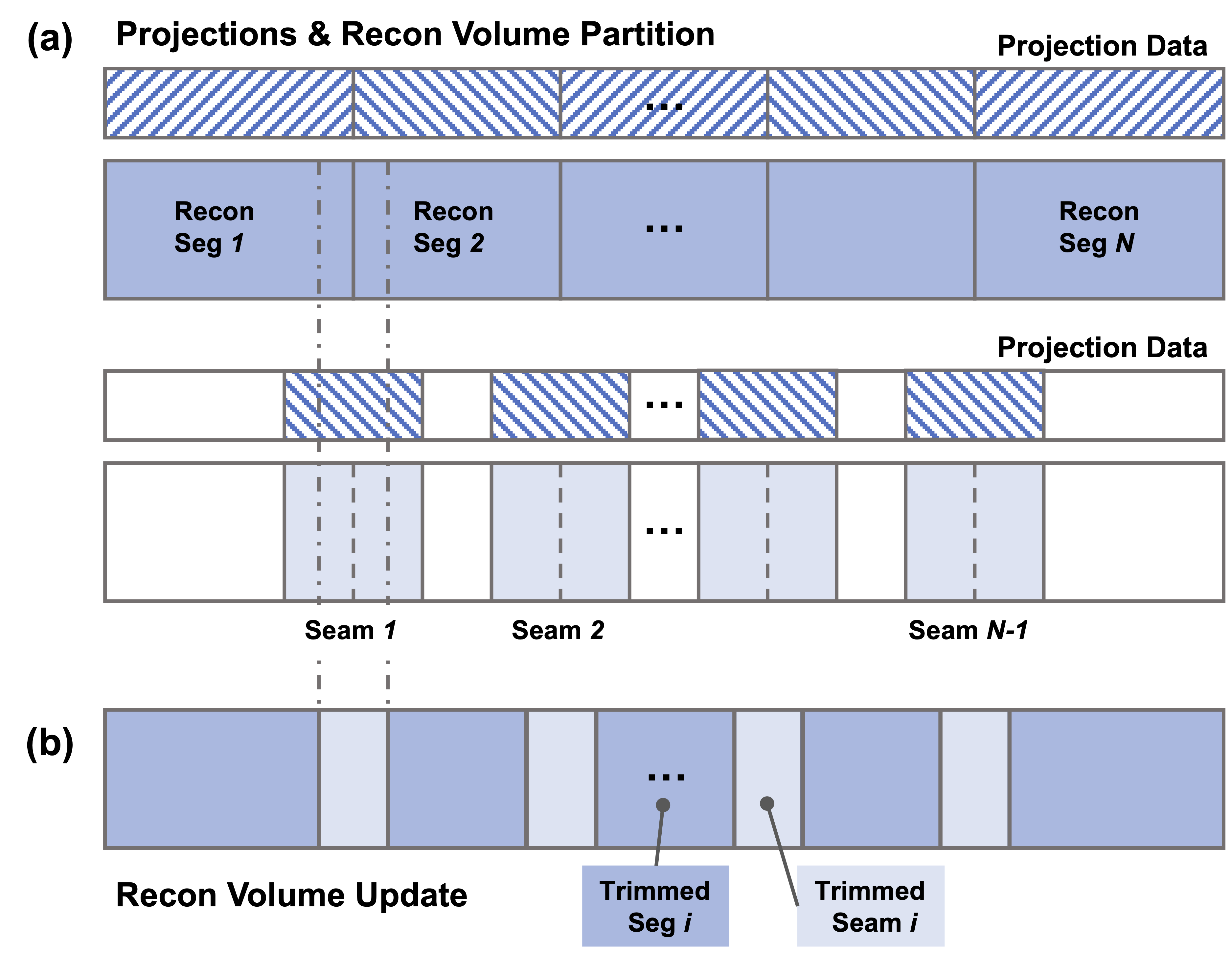}
  \vspace{-3truemm}
  \caption{Interleaved updating for large volume reconstruction. (a) Partitioning the projections and image volume to form a batch of tasks for sub-volume reconstruction, and (b) combining the results in an interleaved pattern with slices at one or both ends trimmed off to ensure data completeness.
  }
  \label{fig:InterleavedUpdate}
  \vspace{-3truemm}
\end{figure}

The size of projection data from a patient scan can be huge, \emph{e.g.}, \(1,536\text{~columns}\times128\text{~rows}\times3,392 \text{~views}\times5\text{~channels}\), overwhelming the GPU memory during direct reconstruction. 
We further use an interleaved updating technique to divide the large-volume reconstruction task into a batch of mini-jobs in smaller size by partitioning the projection data and reconstruction volume into different segments as illustrated in Fig.~\ref{fig:InterleavedUpdate}. The volume along with the associated projection data is partitioned into $N$ segments, with each volume segment and its corresponding projection data segment being geometrically aligned. To ensure the data completeness, sub-volumes at the seams are also extracted with their corresponding projection data (about 1.5 to 2 rotations from the helical scan). The volume segments and seams form $2N-1$ mini-reconstruction tasks, which are assigned to multiple GPUs for parallel computing or can be processed sequentially with a single GPU. The resultant sub-volumes are combined in an interleaved pattern, with a few slices at one or both ends trimmed off to ensure data completeness of the resting volume, forming a complete large volume reconstruction update as shown in Fig.~\ref{fig:InterleavedUpdate} (b).

\section{Experiments and Results}
\subsection{Implementation and Experimental Setup}
\textbf{Training Details.} Our VSR-Net and FRCAN are implemented on PyTorch and trained with the Adam optimizer on a single NVIDIA V100 GPU. The learning rate for VSR-Net is initially \(2\times10^{-4}\) and decayed by 0.95 every epoch. The total number of epochs is 60 with a batch size of 32. The learning rate for FRCAN is initially \(1\times10^{-4}\) and decayed by 0.6 every epoch, with a total of 10 epochs and a batch size of 32. The VSR-Net is trained on the synthetic dataset described in Sub-sec.~\ref{sec:VSR-Net_training}, while FRCAN is trained on real patient data as described in Sub-sec.~\ref{sec:FRCAN_training}.

\textbf{Reconstruction Details.} We use the ASTRA Toolbox~\cite{van2016fast} for GPU-based forward and backward projection operations. The patient data are reconstructed on a cluster node using four NVIDIA V100 GPUs for parallel computation (parallel sub-volume reconstruction and patch processing), and other data are reconstructed on a server with a single RTX A5000 GPU. 

\textbf{Experimental Setup.} \emph{First}, we demonstrate the in-domain capability of our DIR method on synthetic single channel CT data. The testing volume is generated from the AAPM dataset following a similar simulation protocol but from different patients. \emph{Then}, we demonstrate the enhanced generalization on out-of-domain data with our DIR compared to the conventional post-processing with VSR-Net. We use phantom data scanned from a micro-PCCT system for out-of-domain testing. \emph{Finally}, we validate our whole PCCT reconstruction workflow (DIR followed by texture appearance tuning) on new real patient data acquired on the MARS Extremity scanner. Ratings from radiologists on diagnostic value are used to assess the effectiveness of our method.

\subsection{\color{subsectioncolor}In-domain Simulation Study}
\begin{figure*}
\vspace{-5truemm}
  \centering
  \includegraphics[width=0.95\linewidth]{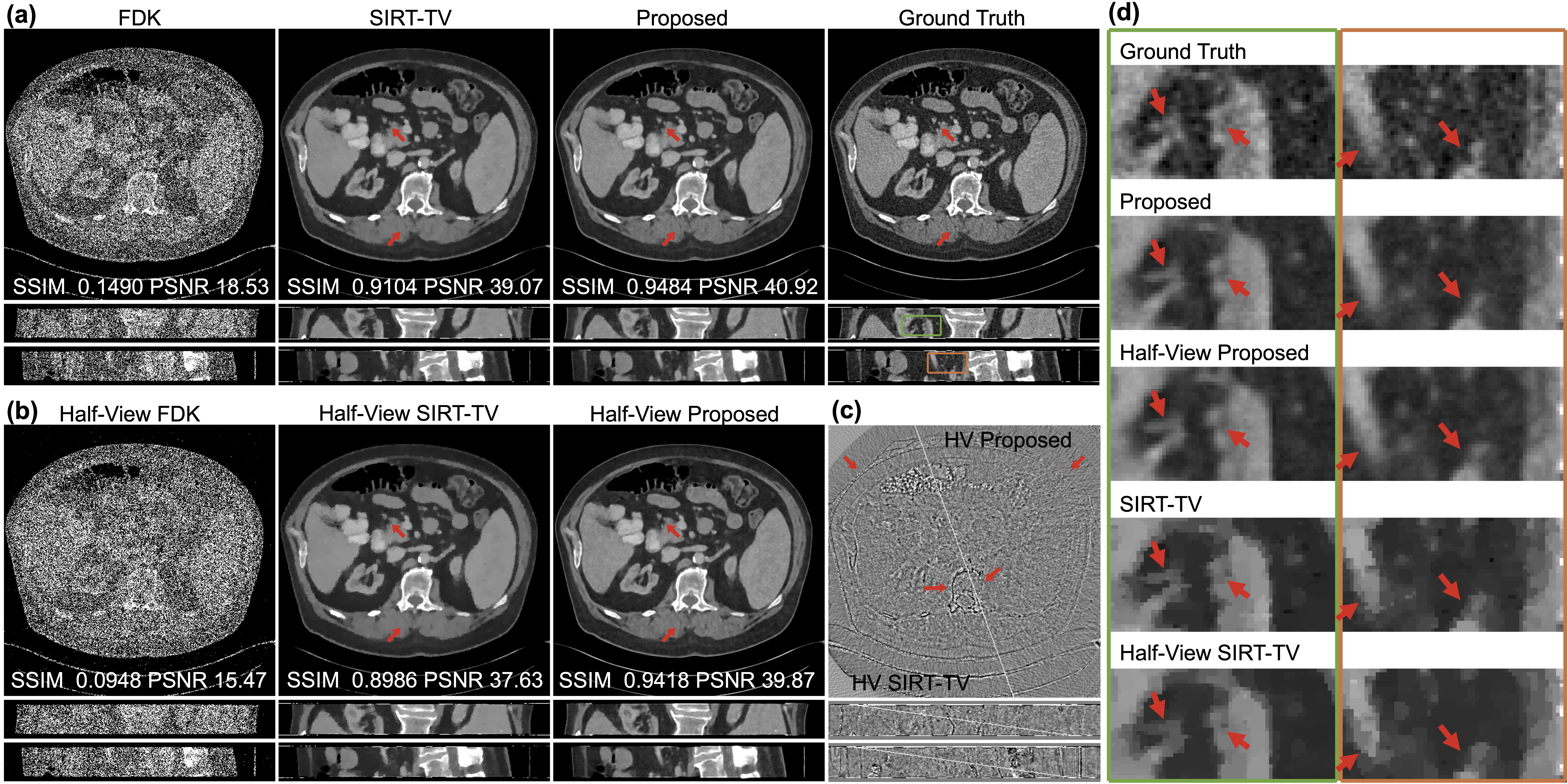}
  \vspace{-3truemm}
  \caption{Representative images reconstructed using the competing methods on simulated data. (a) The full-view reconstructions with FDK, SIRT-TV, and our method displayed against the ground truth, including exemplary axial, coronal, and sagittal views from top to bottom; (b) the reconstructions from halved views; (c) error map of half-view reconstructions against ground truth for the SIRT-TV (left half) and the proposed method (right half); and (d) magnified regions from the coronal and sagittal views as indicated by the green and orange boxes respectively and displayed in the descent order of image sharpness and structural fidelity: ground truth, our full-view and half-view reconstructions, and full-view and half-view reconstructions with SIRT-TV from top to bottom. The display window settings are W/L:400/50 HU for images and W/L:200/0 HU for error maps. The red arrows highlight the structural details that are recovered for our methods but challenging for SIRT-TV, \emph{e.g.}, resulting in loss of resolution as indicated in (c) and a blotchy and cartoonish appearance as shown in (d).}
  \label{fig:SimResult}
  \vspace{-5truemm}
\end{figure*}

\color{black}{
We first evaluate our DIR method on simulated in-domain cone-beam CT data. Specifically, a numerical flat panel detector consists of $1,536\times128$ pixels with \(0.11 mm\) pitch. The source-to-detector distance and the source-to-isocenter distance are \(949 mm\) and \(625 mm\), respectively. Over a full scan, 373 projections are evenly collected, and the number of incident photons per detector element is set to 16,000 in an empty scan. Aside from Poisson noise, we did not include Compton scattering or other physical effects in this simulation, to remain consistent with the training data simulation conditions. The reconstruction volume is set to $420 \times 420 \times 60$ isotropic voxels ($0.2^3 mm^3$).
The SIRT reconstruction from clean projection data with 500 iterations serves as the ground truth. The standard FDK reconstruction from noisy projection data reveals the severity of image noise and artifacts. 

Our proposed method is compared against the anisotropic TV~\cite{shamouilian2021fast} regularized SIRT reconstruction (SIRT-TV) in both full-view and half-view scenarios~\cite{chen2013limited}.
We apply DIR to half-finished SIRT reconstructions with settings of $\mu = 0.01, \beta=0.5,\gamma=0.8$ and $\mu=0.015,\beta=0.5, \gamma=0.8$ in the full-view and half-view cases respectively, and we use 10 gradient descent steps per iteration in both the cases.
Representative full-view and half-view reconstructions are displayed in Fig.~\ref{fig:SimResult}, showing the superior performance of our method despite altered acquisition conditions. 
The fine details indicated by the red arrows are successfully restored with our methods for both full-view and half-view cases while missing structures or distortions are observed with the SIRT-TV particularly in the half-view scenario. Additionally, unnatural waxiness is also observed in the zoomed-in regions of SIRT-TV results. Moreover, our half-view reconstruction scores surpass those of the full-view reconstruction with the conventional method in terms of structural similarity index metric (SSIM) and peak signal-to-noise ratio (PSNR), demonstrating the superiority of our method. More importantly, our method demonstrates impressive stable performance despite a significant change in acquisition condition from full-view to half-view ($<1.0\%$ loss in SSIM and $<4.0\% $ loss in PSNR), which is even more robust than the classic SIRT-TV.}

\subsection{Out-of-domain Phantom Studies}
\begin{figure*}
\vspace{-3truemm}
  \centering
  \includegraphics[width=0.95\linewidth]{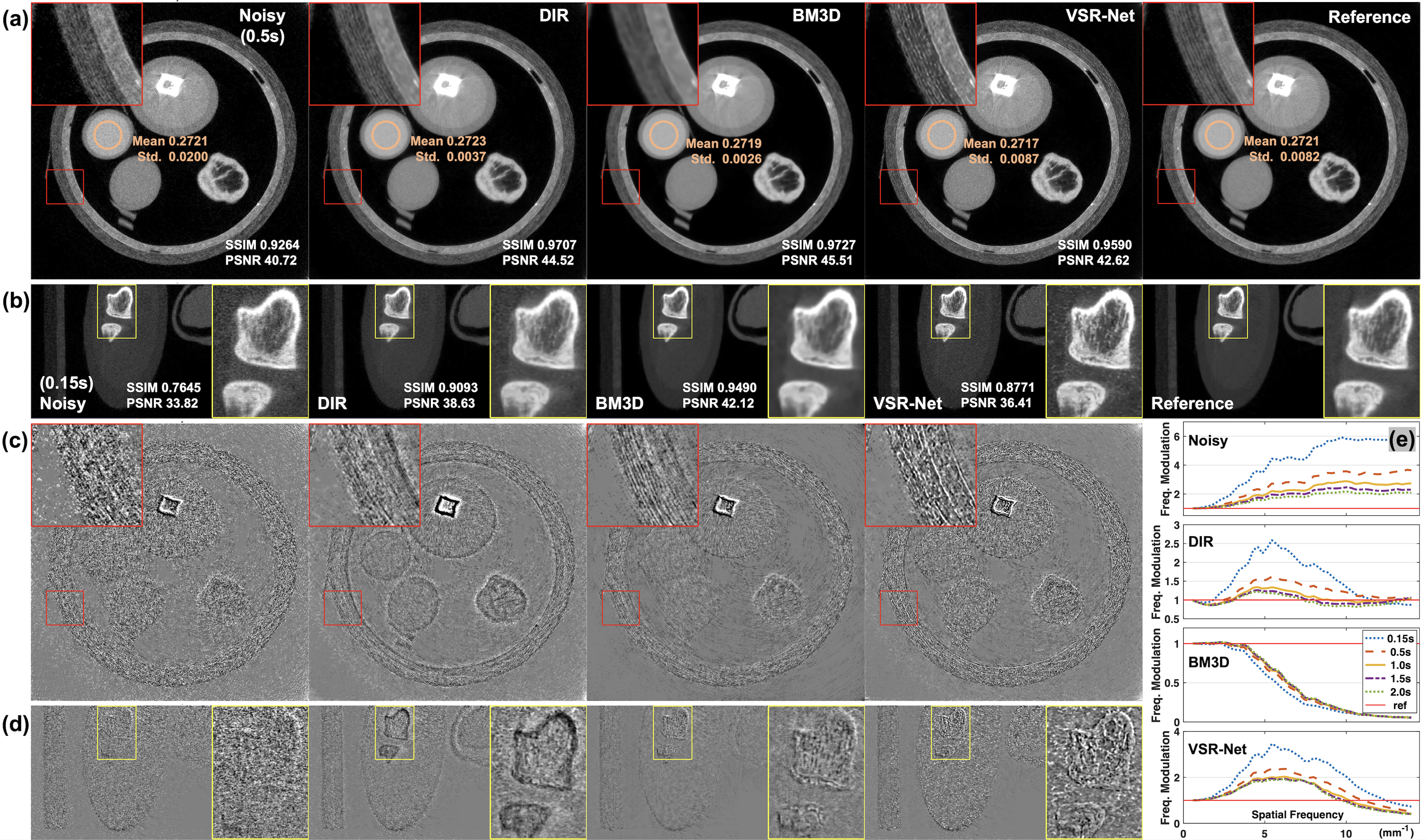}
  \vspace{-3truemm}
  \caption{Generalization on the PCCT phantom scan that is out-of-domain. Comparisons between noisy input, DIR, BM3D, and VSR-Net against the long-exposure reference (5.0 seconds). (a) Axial slices from the dataset with 0.5-second exposure; (b) Sagittal slices from the dataset with 0.15-second exposure; (c) and (d) Difference images for (a) and (b) with respect to the reference, revealing resolution loss in BM3D (particularly in tape details) and falsely generated structures by VSR-Net. In contrast, DIR results exhibit no noticeable loss of resolution or structures, apart from a slight bone value shift in the extreme-case reconstruction shown in (d); and (e) Radial profiles of spatial frequency modulation functions, calculated by azimuthally averaging the normalized Fourier spectra of (a) with respect to the spectrum of the reference, further confirming the resolution loss in BM3D and demonstrating the improvement of DIR over VSR-Net with significantly dampened spectrum deviation. The display window is [0, 0.45] for axial view and [0, 1.05] for sagittal view while the corresponding windows for difference images are [-0.05, 0.05] and [-0.15, 0.15], respectively (unit: cm$^{-1}$). The mean and standard deviation values of a flat water region are listed along with the SSIM and PSNR values.}
  \label{fig:PhantomStudy}
  \vspace{-5truemm}
\end{figure*}

To demonstrate the generalizability, we further test our DIR method on phantom data with totally different structures as shown in Fig.~\ref{fig:PhantomStudy}. The single-channel helical scan data are acquired on our custom-built micro-CT system equipped with a PCD (ADVACAM WidePIX1x5, Prague, Czech Republic) at 80\(kVp\). We collect the data at 6 different dose levels by adjusting the exposure time per projection (0.15, 0.5, 1.0, 1.5, 2.0, and 5.0 seconds). The volumes of $979\times979\times610$ isotropic voxels ($35^3\mu m^3$) are reconstructed using 250 SIRT iterations, serving as noisy inputs (exposure time of 0.15 to 2.0 seconds) and clean reference (5.0-second exposure time). Post-processing with the latest BM3D~\cite{makinen2020collaborative} and with VSR-Net (the same one used in our DIR) are the baselines for our DIR method. The standard deviation parameters for the BM3D method are determined by measuring the standard deviation of values in a water region after normalized with its mean. For the DIR method, we use the half-finished reconstruction as the structural prior (60 and 40 SIRT iterations at the scale of 0.637 and 1 respectively), and refine it with 36 DIR iterations (3 gradient descent steps per iteration, $\mu=0.03$, $\beta=0.5$, $\gamma=0.8$). 

\begin{figure*}
  \vspace{-3truemm}
    \centering
    \includegraphics[width=0.9\linewidth]{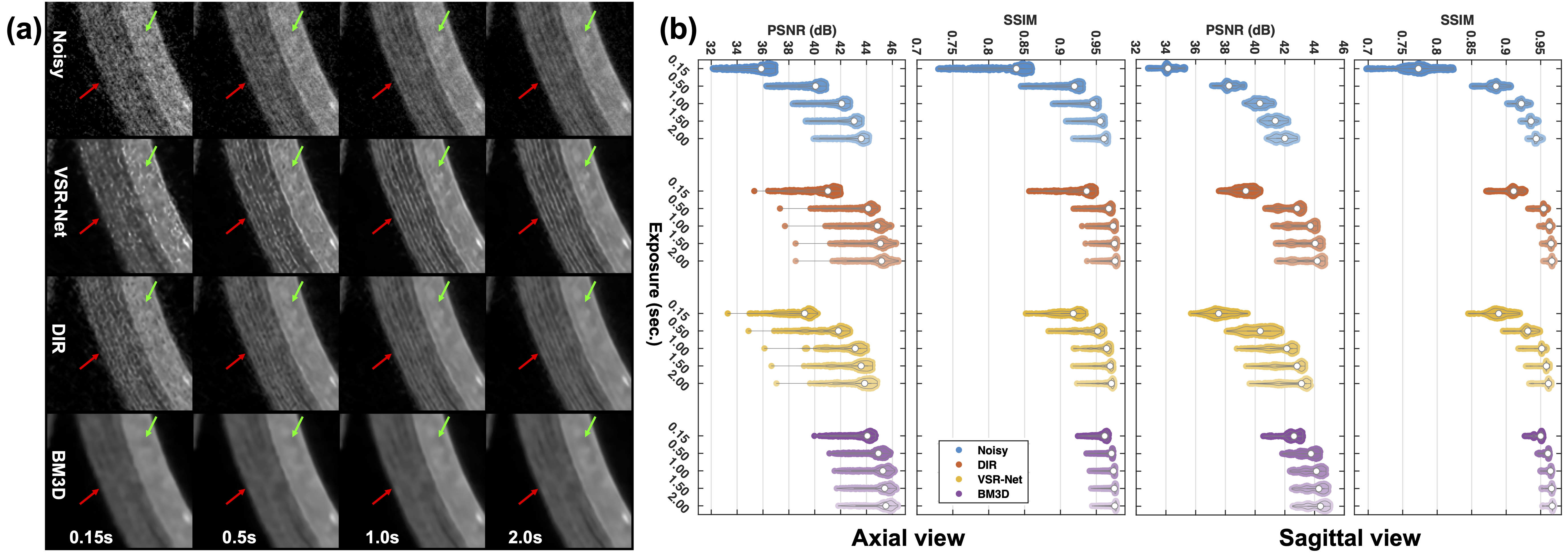}
    \vspace{-3truemm}
    \caption{Zoomed comparison on the phantom scan under different noise levels accompanied by quantitative image quality evaluation. (a)Same magnified region as that in Fig.~\ref{fig:PhantomStudy}(a) but from the datasets with exposures of 0.15, 0.5, 1.0, and 2.0 seconds respectively; and (b)violin plots of distributions of the PSNR and SSIM values for the axial and sagittal slices.}
    \label{fig:PhantomStudy2}
    \vspace{-5truemm}
  \end{figure*}

Figures~\ref{fig:PhantomStudy}(a) to (d) compare the results with 0.5 and 0.15 seconds exposure from  axial  and sagittal views against the reference, respectively. Figure~\ref{fig:PhantomStudy}(e) displays the frequency modulation curves of axial slices from \ref{fig:PhantomStudy}(a) related to the reference. 
Figure~\ref{fig:PhantomStudy2}(a) illustrates a zoomed-in view of a surgical tape under various exposure times and reconstruction methods. For each combination, we computed the SSIM and PSNR for every slice (either from the axial or sagittal view) from the volume against the reference for quantitative comparison. The resulting SSIM and PSNR distributions are presented in Fig.~\ref{fig:PhantomStudy2}(b) using violin plots with boxplot overlay. In each violin, the center thick gray line represents the interquartile range, and the large white dot indicates the median.
Our method consistently demonstrates improved image quality across different acquisition conditions in terms of both qualitative and quantitative measures, suggesting good generalizability on out-of-domain structures. In contrast, the adverse effects of domain shift are clearly presented in VSR-Net results, showing artifacts and different appearance and intensity from that of the reference despite the enhanced structure visibility, \emph{e.g.}, structural errors revealed in Figs.~\ref{fig:PhantomStudy}(c) and (d), and the tape structure and dots pointed by the red and green arrows in Fig.~\ref{fig:PhantomStudy2}(a). The BM3D method is structurally agnostic and intrinsically offers better generalization than deep learning methods as reflected by the high PSNR and SSIM scores in Fig.~\ref{fig:PhantomStudy2}(b). However, it suffers from resolution loss as observed in the difference image in Fig.~\ref{fig:PhantomStudy}(c) and a significantly dampened spectrum in the middle and high frequencies in Fig.~\ref{fig:PhantomStudy2}(e), compromising subtle details. This suggests the importance of task-relevant metrics and underlines the need for radiologists' evaluation in the medical imaging field.

\subsection{Retrospective Patient Study}
Patients aged 21 years and above referred from the fracture clinic are recruited for the clinical trial (Ethics approval:18/STH/221/AM01, Health and Disability Ethics Committee, New Zealand). The spectral image volumes of patient wrist are acquired using the MARS Extremity PCCT scanner with scanning settings shown in Table~\ref{Table:ScanSettings}. 

\begin{table}
  \caption{Scanner Settings for Patient Study.}
  \vspace{-3truemm}
  \footnotesize
  
  \begin{center}
    \begin{tabular*}{0.95\linewidth}{l@{\extracolsep{\fill}}c}
      \toprule
      Parameters & MARS Extremity 5X120 Settings \\
      \midrule
      Tube Setting    & 118\(kVp\), 28\(\mu A\)  \\
      Exposure    & 160\(ms\) per projection, helical scan \\
      CTDI$_{\text{vol}}$ & 7.87\(mGy\)\\
      Energy Bins    &  \(7-40\), \(40-50\), \(50-60 \), \(60-70\), \(70-118\), \(keV\)\\
      Recon. Voxel   & Isotropic \(90\times90\times90 \mu m^3\) voxel \\ 
      Recon. Method & A customized polychromatic iterative method.~\cite{de2017mars}\\
      \bottomrule
    \end{tabular*}
    
  \end{center}
  
  \label{Table:ScanSettings}
  \vspace{-7truemm}
\end{table}

\subsubsection{Experiment Setup}
The images of 8 patients who provided written consents are reconstructed using the commercial algorithm from a full-view dataset and our proposed deep learning method (illustrated as Fig.~\ref{fig:Diagram}) from a half-view dataset respectively, and then evaluated by three independent double-blinded radiologists (SG, AB, AL) using the rating scale defined in Table~\ref{Table:Scale} regarding whether diagnostic image quality is achieved or not~\cite{bath2007visual}. In our method, DIR is applied to the structural prior (Sub-sec.~\ref{sec:StructurePrior}, obtained with 80 and 80 SIRT iterations at the scale of 0.5 and 1 respectively) for 30 iterations (3 gradient descent steps per iteration, \(\mu=0.03,\beta=0.5,\gamma=0.8\)) for the reconstruction from data in each bin, then the combined multi-channel volume (\(1,200^3\times5\)) are processed with RFCAN in a slice-by-slice manner for value alignment and texture enhancement (Sub-sec.~\ref{sec:TextureTuning}), and the number of SIRT iterations is set to 30 and the mixing ratio to \(0.75\colon0.25\) to accommodate radiologists' preference on image sharpness and noise characteristic. 

\begin{table}[htbp]
\vspace{-3truemm}
  \caption{Grading scale for image quality assessment.}
  \vspace{-3truemm}
  \footnotesize
  \begin{center}
    \begin{tabular*}{0.45\textwidth}{l@{\extracolsep{\fill}}cc}
      \toprule
      -2 & Confident that the diagnostic criteria is not fulfilled; \\
      -1 & Somewhat confident that the criteria is not fulfilled;   \\
      0  & Indecisive whether the criteria is fulfilled or not;     \\
      +1 & Somewhat confident that the criteria is  fulfilled;      \\
      +2 & Confident that the criteria is fulfilled.                \\
      \bottomrule
    \end{tabular*}
  \end{center}
  \label{Table:Scale}
  \vspace{-5truemm}
\end{table}

\begin{figure}
  \centering
  \includegraphics[width=0.99\linewidth]{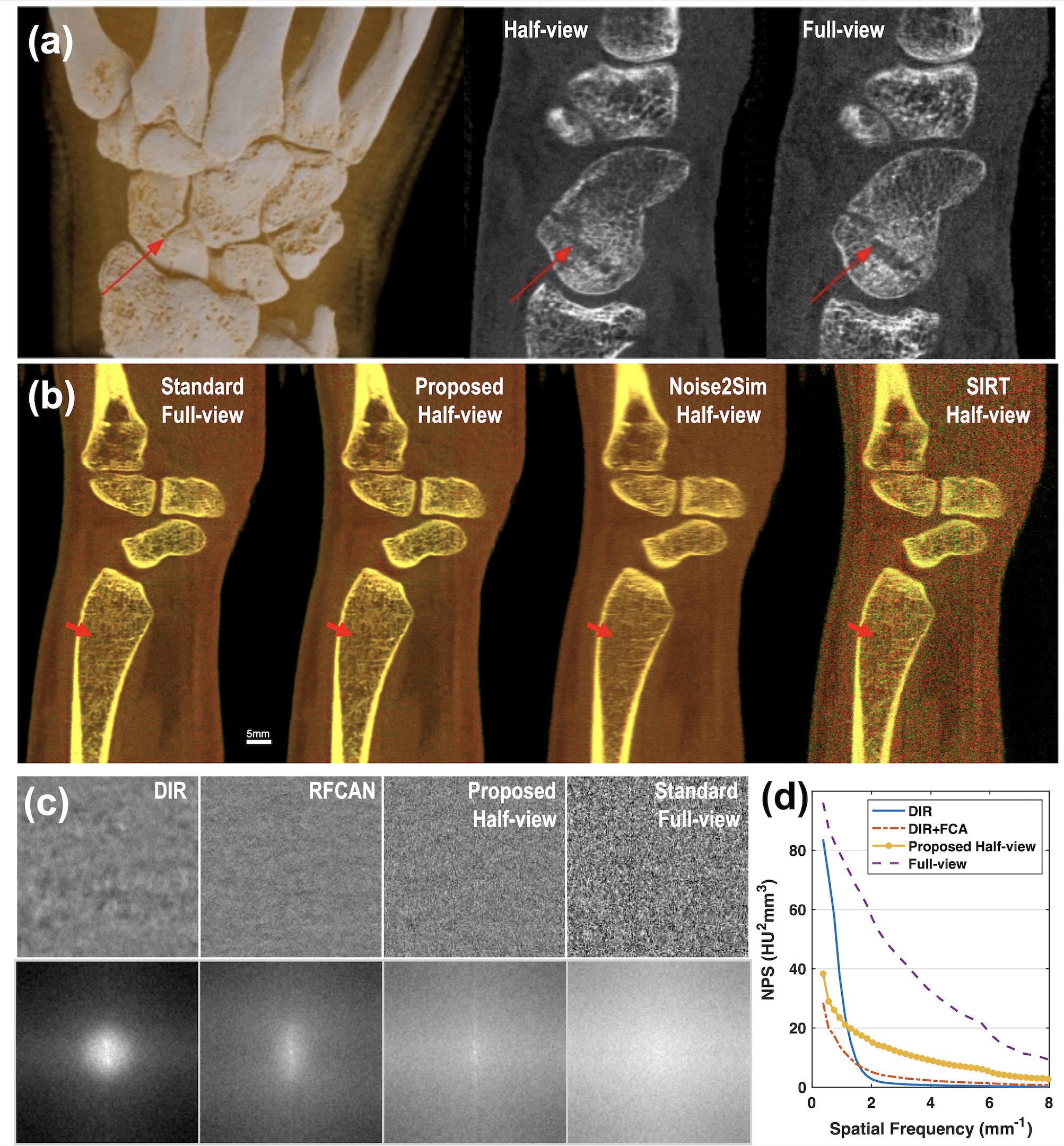}
  \vspace{-7truemm}
  \caption{{Sagittal view of a wrist joint reconstructed using the standard and proposed methods respectively}. (a) From left to right are 3D rendering of the standard reconstruction, half-view and full-view images in the channel 7-40keV (manually aligned to show the same region, displayed in W/L:0.72/0.5 cm$^{-1}$), where the arrow points to a scaphoid fracture; (b) a color visualization of our three-channel reconstruction via linear blending~\cite{li2024contrast} in reference to the standard full-view result and Noise2Sim half-view result. Our result demonstrates high fidelity in both spectral values (same color tone and brightness as the full-view) and spatial structures (sharp and accurate fine details as pointed by the red arrow); (c) zoomed-in views of noise texture from a flat tissue region in our half-view reconstruction (7-40keV channel) illustrate the texture changes across different stages; and (d) the corresponding NPS curves demonstrate that, despite differences in noise amplitude, our final result closely matches the reference in spectral shape.}
  \label{Fig:RecResultOfPatient}
  \vspace{-6truemm}
\end{figure}

The radiologists are randomly presented with 500 images from each patient (three energy bins \(7-40keV\), \(50-60keV\) and \(70-118 keV\)) in the axial, coronal and sagittal formats. The sagittally reformatted images reconstructed using both methods along with 3D rendered images are shown in Fig.~\ref{Fig:RecResultOfPatient}(a). 
The image metrics are based on the ``European guidelines on quality criteria for CT" for bones and joints~\cite{carmichael1996european}, including the visibility and sharpness of the cortical and trabecular bone, the adequacy in soft tissue contrast for the visualization of tendons, muscle and ligaments, as well as image noise (quantum noise) and artifacts.

Additionally, we compare our results with the state-of-the-art results obtained by applying the self-supervised denoising method Noise2Sim~\cite{niu2022noise} to the multi-channel reconstructions after 320 SIRT iterations. Despite significant enhancement over SIRT reconstruction from the half-view dataset, Noise2Sim results demonstrate insufficient image quality (suffering from image blur and losing fine structures) as shown in Fig.~\ref{Fig:RecResultOfPatient}(b). Hence, they are excluded from the reader study.

\subsubsection{Data Analysis}
For quantitative assessment, regions of interest (ROIs), each with $\sim250$ voxels, are drawn in the flexor carpi radialis tendon and adjacent subcutaneous fat regions in the patient images. The mean and standard deviation of linear attenuation coefficients in the ROIs are used to calculate the signal-to-noise ratio (SNR) in soft tissue regions and contrast-to-noise ratio (CNR) between soft tissue and fat in the ROIs. SNR and CNR values associated with both methods are compared over all patients' datasets.
For subjective evaluation, overall radiologists' assessment grades for seven image quality measures from both methods are summarized as a frequency table. Then, all three radiologists' overall and combined ratings are compared with descriptive statistics. The hypothesis of no significant difference between the two methods is tested in the Wilcoxon signed-rank test.

Image grades from both methods in terms of each image quality metric are also converted into visual grading characteristics (VGC) points as described in~\cite{bath2007visual}. Hence, with the current 5 image grading criteria, 4 VGC points are obtained, and 0 as the origin and 1 as the maximum value are added as well~\cite{bath2007visual}. The combined VGC points (using grades from all three radiologists) of seven image quality measures are calculated, and the empirical area under the curve (AUC$_{VGC}$) is compared. The statistical significance of the mean AUC$_{VGC}$ of seven image quality measures is analyzed through one-sample t-testing against a hypothetical AUC$_{VGC}$ value of 0.5 (the value 0.5 corresponding to equal/comparable image quality for the two imaging methods in comparison). VGC points are also obtained by combining all seven image quality measures for both methods. Full-view vs Half-view empirical AUC$_{VGC}$ values and their 95\% confidence intervals are obtained for each radiologist and the combined scores. The statistical significance of AUC$_{VGC}$ and its 95\% confidence interval for each radiologist (56 samples) and all radiologists (168 samples) are interpreted according to the method described in~\cite{hanley1982meaning}. The statistical analysis is presented using GraphPad Prism 9.2.0 at a significance level of 95\%. Finally, the inter-rater agreements between radiologists are evaluated with quadratically weighted kappa statistics~\cite{fleiss1969large}.

\subsubsection{Image Comparison}
Figure~\ref{Fig:RecResultOfPatient}(a) displays a 3D rendering of a patient's wrist with a scaphoid fracture, along with oblique slices from our half-view reconstruction and the standard full-view reconstruction, highlighting the lesion. The fracture is clearly presented in two images, showing comparable contrast and noise textures. Note that noticeable structural differences are due to slight mismatches in slice location or viewing angle introduced by manual alignment. A color visualization of the three-channel spectral images via linear blending~\cite{li2024contrast} is shown in Fig.~\ref{Fig:RecResultOfPatient}(b). The consistent color tone and brightness compared to the standard result demonstrate the high fidelity of our result in spectral values. Fine structural details are also well-preserved, as indicated by the red arrow. In contrast, the Noise2Sim result exhibits structure loss, while the conventional SIRT result suffers from significant noise and contrast issues. To further evaluate the noise characteristics, Fig.~\ref{Fig:RecResultOfPatient}(c) presents zoomed-in noise textures from a flat tissue region at different stages of our reconstruction, alongside the standard reference and their corresponding Fourier spectra. The corresponding NPS curves are plotted in Fig.~\ref{Fig:RecResultOfPatient} (d). As revealed in the figures, the DIR result contains major low-frequency noise components. These are progressively replaced by higher-frequency components through RFCAN post-processing, with the final reconstruction further enhancing high frequencies and well resembling the texture of the standard reference, despite exhibiting a lower overall noise amplitude.

\subsubsection{Statistical Results}
\begin{figure}
  \centering
  \includegraphics[width=0.85\linewidth]{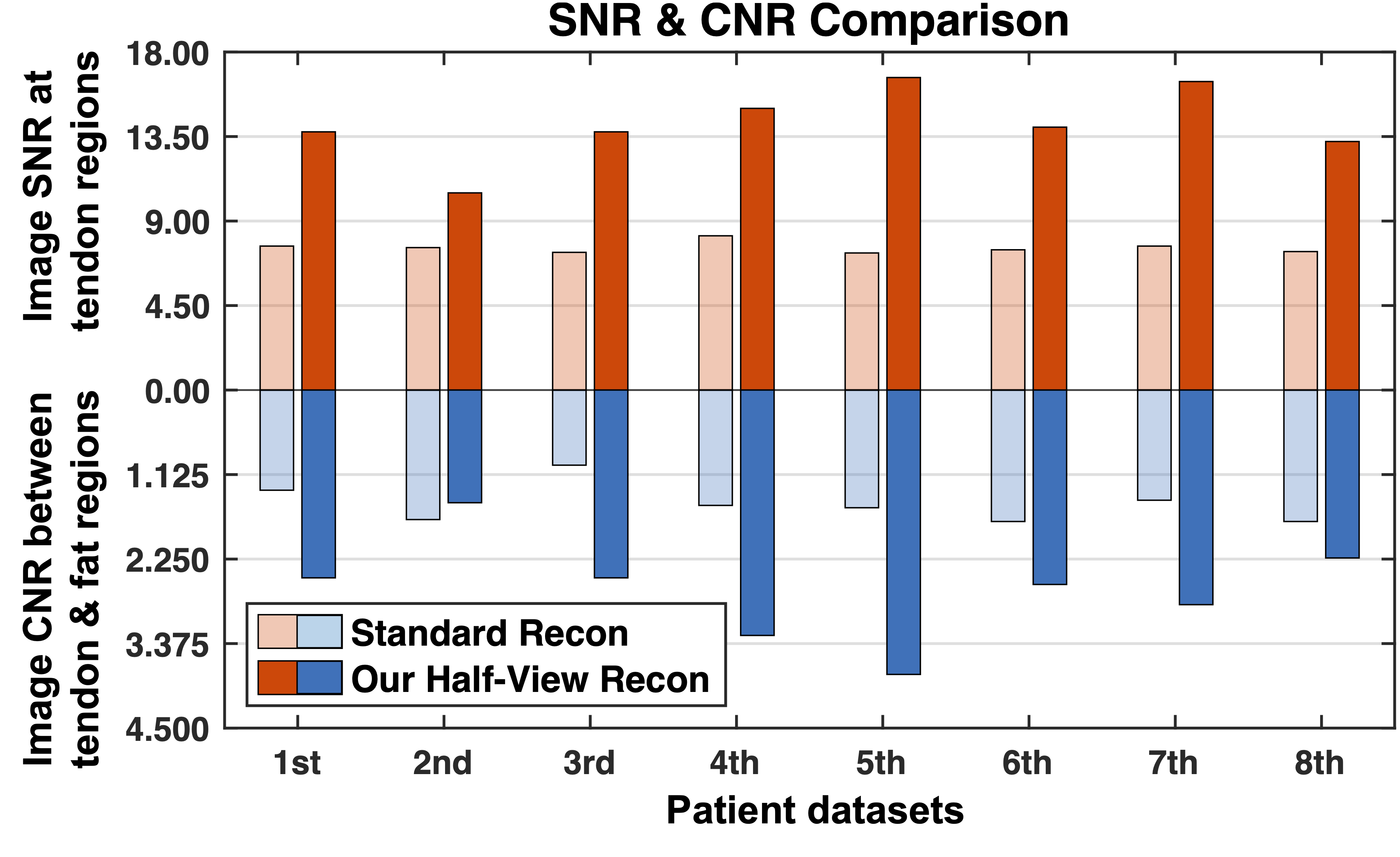}
  \vspace{-3truemm}
\caption{Quantitative evaluation of our half-view reconstruction against the standard commercial reconstruction from the full-view dataset.}
  \label{fig: final graph A}
  \vspace{-3truemm}
\end{figure}

\begin{figure}
    \centering
    \includegraphics[width=0.65\linewidth]{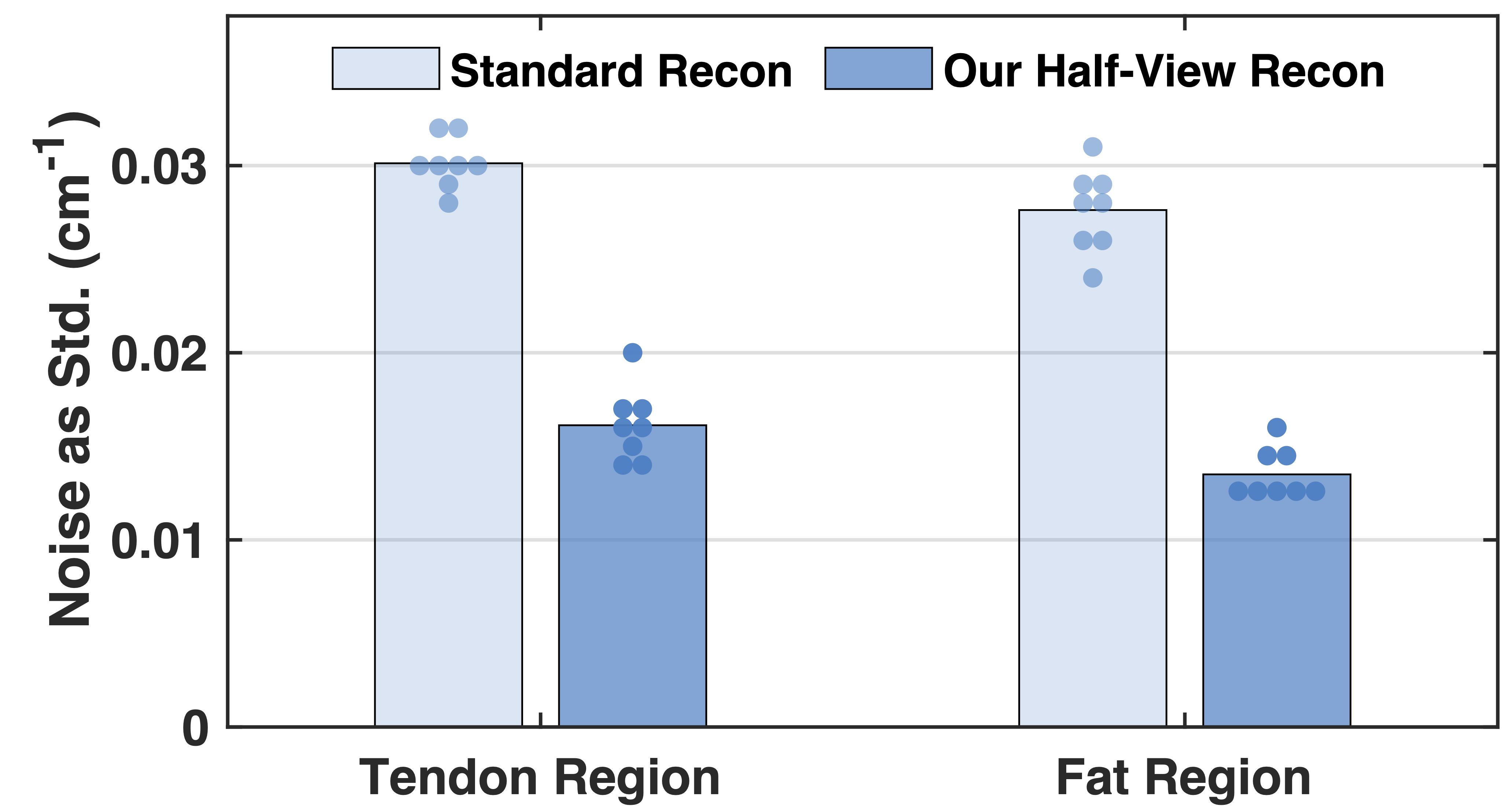}
    \vspace{-3truemm}
  \caption{Distribution of the noise across patients measured as standard deviations of pixel values within the regions for CNR/SNR calculation.}
    \label{fig:NoiseDistr}
    \vspace{-5truemm}
  \end{figure}

SNR in soft tissue regions and CNR between soft tissue and fat are compared in Fig.~\ref{fig: final graph A}, where the bar charts illustrate that for all the patients, SNR and CNR in the images obtained with the proposed half-view reconstruction method are higher than that in the clinical benchmark images reconstructed using the standard commercial method from the full dataset, except for the second one whose images showed quite comparable CNRs. The distributions of image noise across patients shown in Fig.~\ref{fig:NoiseDistr} further indicates that the proposed method consistently yields significantly lower noise levels compared to the standard method, despite utilizing halved radiation dose.

\begin{table}
\vspace{-3truemm}
  \caption{Descriptive statistics of the radiologists' ratings. 
  }
  \vspace{-3truemm}
  \footnotesize
  \begin{center}
    \begin{tabular*}{0.95\linewidth}{l@{\extracolsep{\fill}}rccr}
      \toprule
      Methods & ~Raters & Median $\uparrow$ & IQR & Mean $\uparrow$ (Std.) \\
      \midrule
      Full    & RD1    & 1   &0  & 0.875 (0.740)            \\
      Full    & RD2    & 1   &2  & 1.107 (0.966)          \\
      Full    & RD3    & $-$1&3  & $-$0.589 (1.247)       \\
      Full    & COM    & 1   &1  & 0.464 (1.252)          \\
      Half     & RD1    & 1  &2  & 1.054 (0.862)         \\
      Half     & RD2    & 2  &2  & 1.179 (1.011)          \\
      Half     & RD3    & 0  &3  & $-$0.357 (1.354)       \\
      Half     & COM    & 1  &2  & 0.625 (1.293)         \\
      Overall & RD1    & 1  &1.5 & 0.964 (0.804)          \\
      Overall & RD2    & 1  &2  & 1.143 (0.985)           \\
      Overall & RD3    & $-$1&3 & $-$0.473 (1.301)       \\
      Overall & COM    & 1   &2  & 0.545 (1.273)          \\
      \bottomrule
    \end{tabular*}

    \vspace{1ex}
    {\raggedright \footnotesize \hspace{0.025\textwidth}\parbox{0.9\linewidth}{IQR: interquartile range; Full, Half: The standard commercial full-view reconstruction and our half-view reconstruction; Overall: Ratings by combining two methods; RD1, RD2, RD3, COM: Three radiologists and their combined ratings.} \par}
  \end{center}
  \label{Table:DesSta}
\end{table}

\begin{table}
\vspace{-5truemm}
  \caption{Hypothesis testing (Half-view vs Full view) in terms of the Wilcoxon signed rank.}
  \vspace{-3truemm}
  \footnotesize
  \begin{center}
    \begin{tabular*}{0.95\linewidth}{l@{\extracolsep{\fill}}ccc}
      \toprule
      Raters & ~\# of Pairs & \# of Ties & \(p\)-Value \\
      \midrule
      RD1    & 56                       & 22                                                  & 0.2734  \\
      RD2    & 56                       & 44                                                & \(0.3877\) \\
      RD3    & 56                       & 38                                                & \(0.0355\) \\
      COM    & 168                      & 104                                                 & \(0.0166\) \\
      \bottomrule
    \end{tabular*}

      \vspace{1ex}
      {\raggedright \footnotesize \hspace{0.05\linewidth}RD1, RD2, RD3, COM: Three radiologists' and combined ratings. \par}
  \end{center}
  \label{Table:HypTest}
  \vspace{-7truemm}
\end{table}

More importantly, the overall confidence ratings of diagnostic image quality with seven criteria are compared in Table~\ref{Table:DesSta}.  
The table shows significantly better mean and median image quality scores with the proposed half-view reconstruction method than the current clinically used reconstruction method from the full-view dataset for all radiologists despite their different scores, indicating a preference for the proposed reconstructions. The median value for the half-view reconstructions is 2 for the second radiologist, which suggests higher confidence in image interpretation. Despite the discrepancy in ratings, the combined median value is clearly positive, indicating the favorable acceptability of our reconstructions. The hypothesis is further tested in the Wilcoxon signed rank test for all three raters and combined ratings in Table~\ref{Table:HypTest}. It shows that the p-value is not statistically significant for the 1st and 2nd radiologists, suggesting no difference in image quality between the two methods. However, the p-values for the 3rd radiologist and the combined results are statistically significant, indicating the image quality from the proposed method is perceived significantly better than the standard commercial image reconstruction from the full-view dataset.

\begin{figure*}
\vspace{-3truemm}
  \centering
  \includegraphics[width=0.9\linewidth]{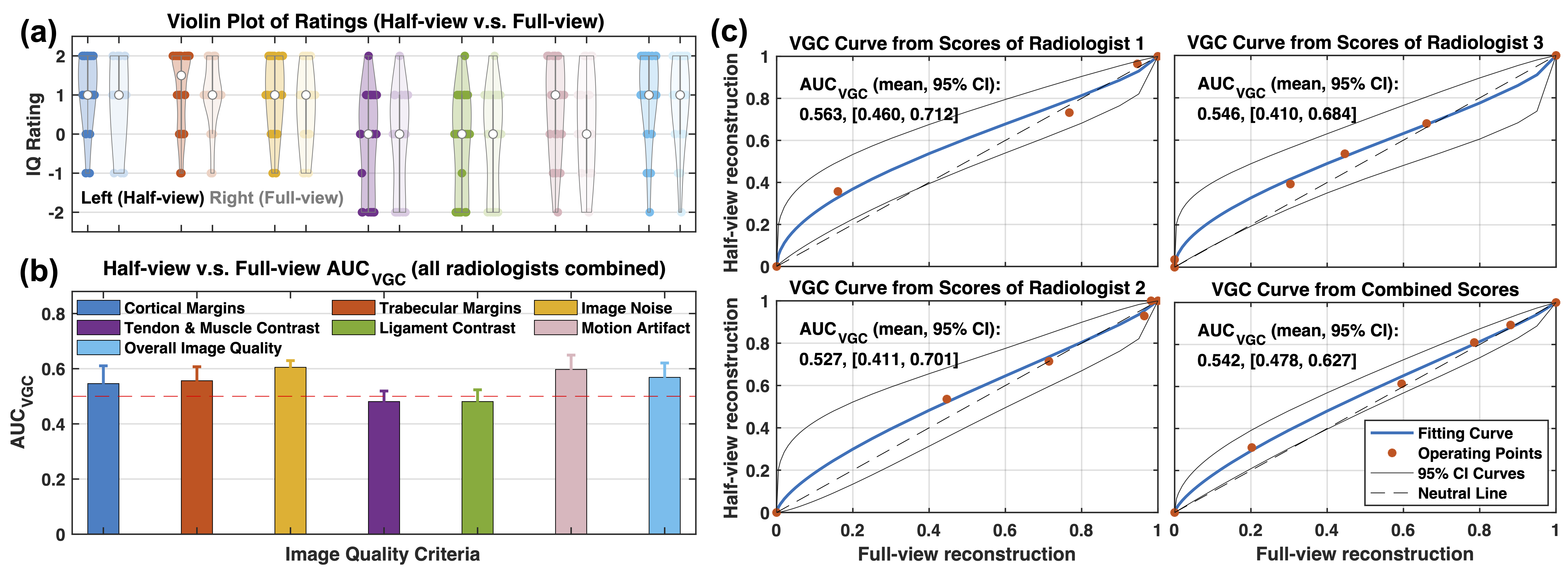}
  \vspace{-3truemm}
  \caption{{Subjective evaluation on seven image quality (IQ) metrics comparing our half-view reconstruction with the clinical full-view reconstruction.} (a) Violin plot of combined radiologists' ratings on the half-view results against the full-view results; (b) Slightly better than 0.5 AUC$_{VGC}$ (the performance neutral threshold) from the proposed method over the conventional method for most image quality metrics; and (c) the half-view versus full-view VGC plots generated by combining all the image quality metrics indicate most VGC points above the diagonal line (the performance neutral line) for all the three radiologists and their combined ratings.}
  \label{fig: final graph B}
  \vspace{-5truemm}
\end{figure*}

The proposed method also performs better when image quality measures are individually evaluated. The mean area under the curve for visual grading characteristics AUC$_{VGC}$ values of the proposed method are consistently higher than 0.5 for five image quality measures evaluated, as shown in Fig.~\ref{fig: final graph B}(b). Similar trends are also reflected in the violin plots in Fig.~\ref{fig: final graph B}(a) as indicated by the better median scores and narrower tails in the low end (less low scores). The mean AUC$_{VGC}$ from the standard method was only slightly better than that of the proposed method in soft tissue contrast differentiation, mainly related to the depiction of ligaments, tendons, and muscles. The VGC points obtained for overall image quality scores from the two competing methods are plotted in Fig.~\ref{fig: final graph B}(c). Fig.~\ref{fig: final graph B}(c) shows that the mean value of AUC$_{VGC}$ for the proposed method is better than 0.5 for all radiologists and combined ratings. However, to show that the AUC$_{VGC}$ is significantly better than 0.5 in the $95\%$ interval sense, more data would be needed. The statistical significance of the mean of AUC$_{VGC}$ of the seven image quality measures is established as well using the one-sample t-test in Table~\ref{Table:HypTestAuC}. As the clinical trials proceed, more datasets may help further strengthen the statistical significance of this comparative study.

Although all the raters preferred the proposed half-view reconstructions, they provide different ratings for the same images, resulting in a lower inter-rater agreement. The agreements between raters are evaluated with weighted kappa statistics in Table~\ref{Table:RadAgreement}. The table shows a slight to fair agreement between radiologists' scores of high significance. Also, the kappa value is higher for radiologists 2 and 3, indicating a higher degree of agreement between these two radiologists.

\begin{table}
\vspace{-3truemm}
  \caption{Hypothesis testing of the mean of AUC$_{VGC}$ of the image quality measures (one-sample t-test).}
  \vspace{-3truemm}
  \footnotesize
  \begin{center}
    \begin{tabular*}{0.99\linewidth}{l@{\extracolsep{\fill}}cccc}
      \toprule
      \# of samples & Mean (Std.) & 95\% CI & \(t\)  & \(p\)-Value   \\
      \midrule
      7 (df=6)  & 0.5479 (0.0503)  & [0.0014, 0.0944] & 2.52 & 0.0454 \\
      \bottomrule
    \end{tabular*}

    \vspace{1ex}
    {\raggedright \footnotesize \hspace{0.08\linewidth}\parbox{0.85\linewidth}{Note that 95\% confidence interval (CI) indicates confidence in discrepant value from the hypothetical mean (0.5).} \par}
  \end{center}
  \label{Table:HypTestAuC}
  \vspace{-3truemm}
\end{table}

\begin{table}
  \caption{Agreement in the combined subjective scores between radiologists (quadratically weighted kappa). }
  \vspace{-3truemm}
  \footnotesize
  \begin{center}
    \begin{tabular*}{0.9\linewidth}{l@{\extracolsep{\fill}}cc}
      \toprule
      Categories & Weighted Kappa & \(p\)-Value \\
      \midrule
      RD1-RD2    & 0.247 & \(0.0042\)       \\
      RD1-RD3    & 0.171 & \(<0.0001\)        \\
      RD2-RD3    & 0.339 & \(<0.0001\)       \\
      \bottomrule
    \end{tabular*}

    \vspace{1ex}
    {\raggedright \footnotesize \hspace{0.1\linewidth}\parbox{0.8\linewidth}{RD1, RD2 and RD3 denote the three radiologists, respectively.} \par}
  \end{center}
  \label{Table:RadAgreement}
  \vspace{-8truemm}
\end{table}

\section{Discussion}
This study targets deep learning-based HR PCCT volumetric reconstruction given insufficient training data. Direct volumetric reconstruction is necessary and advantageous since rebinning to fan-beam geometry could compromise image quality~\cite{noo1999single}, especially due to large gaps and bad pixels in PCDs and free-form scanning with robotic arms~\cite{li2022motion,li2022motion2}. However, volumetric PCCT reconstruction poses GPU memory and computational challenges. We have tackled them with interleaved updating, patch-based refinement, and low-noise prior sharing. Low-level structural similarity has been leveraged in combination with model-based iterative refinement to address the domain gap effectively. Additionally, textural appearance has been fine-tuned to align with the standard reconstruction in the application domain. On the other hand, the patch-based representation mitigates the GPU memory limitation but at a cost of extra computation during inference. For example, the reconstruction volume is partitioned into overlapping patches to minimize checker-board artifacts. Compared to directly processing the entire volume, it becomes the overhead to compute the overlapping portions, and perform procedures of partition and assembly and special geometric self-ensemble processing steps. Fortunately, we may use different patch sizes for training and inference, leveraging the shift-invariant property of convolution networks, and bigger patches can be used to reduce the overhead during inference. With parallel computing on four V100 GPUs, a typical computation time to reconstruct a \(94mm\) patient wrist scan in five energy bins is about 7 hours using our method, whereas the commercial reconstruction time is around 9 to 10 hours on its dedicated hardware.
In comparison with the current commercial reconstruction from the full-view dataset, the major benefits of our approach include halved radiation dose and doubled imaging speed, without compromising image quality. The evaluation methods are classic and double blinded. The involved patient datasets are randomly determined, covering a range of pathological and technical conditions. Therefore, our results strongly suggest a great potential of our approach for clinical PCCT image reconstruction. Further improvements are surely possible, using more advanced network architectures such as the emerging diffusion/score-matching models~\cite{gao2024corediff}. The barriers for adapting the diffusion approach for PCCT include the scarcity of high-quality data, the memory limitation, and the sampling overhead, which are being explored actively.

\begin{figure}
    \centering
    \includegraphics[width=0.95\linewidth]{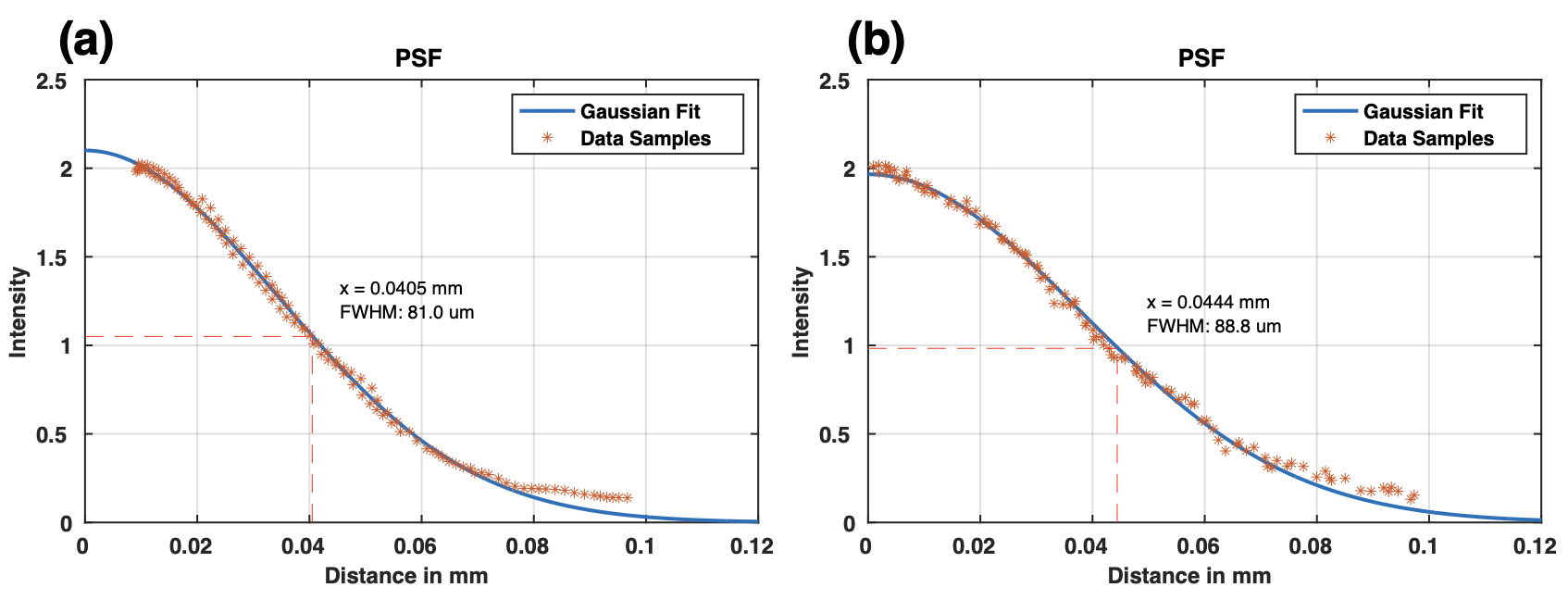}
    \vspace{-3truemm}
  \caption{PSF measurements on a tungsten-wire phantom scan (0.5-second exposure) reconstructed by (a) DIR and (b) SIRT methods. All other settings remained consistent with the phantom study.}
    \label{fig:PSFmeas}
    \vspace{-7truemm}
  \end{figure}

Interestingly, for out-of-domain single-channel real phantom data, significant improvement in both image quality and stability has been made with our DIR approach over the conventional single-pass post-processing method using the same network. For post-processing methods, the domain gap could lead to inappropriate frequency elevation in network processed images and cause undesired artifacts. For example, in Fig.~\ref{fig:PhantomStudy}(e), the spectrum elevation of VSR-Net curves even extends beyond the cutoff frequency, which explains the network's tendency to mistake noise as features and produce high-frequency artifacts. Through iterative feedback and correction, DIR significantly reduces the errors despite its similar elevation pattern to the VSR-Net curves. One limitation is that we might observe some low-frequency bone value shift in the extreme case reconstruction (0.15-second exposure against the 5-second reference, $3\%$ dose) as shown in Fig.~\ref{fig:PhantomStudy}(d), and which is also reflected by a small dent from the DIR frequency modulation profile. This dent can cause low-frequency image contrast change, which explains why the SSIM and PSNR scores of DIR are smaller than those of BM3D. Luckily, no structure or resolution loss is noticed and this issue can be easily remedied with a texture appearance tuning network RFCAN. We further confirmed the resolution change with point spread function (PSF) measurement using a $10\mu m$-diameter tungsten wire phantom. The measured PSF from DIR reconstruction is $81\mu m$ in full width at half maximum (FWHM) while the FWHM from direct SIRT reconstruction reads $88.8\mu m$ as shown in Fig.~\ref{fig:PSFmeas}, suggesting no resolution loss. One interesting observation is that the BM3D method scores the best in terms of SSIM and PSNR despite the loss of fine details, suggesting the necessity of using task-relevant metrics in clinical applications. We also underline that our major aim is to demonstrate the improved generalizability with DIR over VSR-Net in this experiment, rather than to compete with BM3D in these scores. While we already tied scores with BM3D in less noisy cases, superior results can be expected if we adopt an RFCAN or retrain the VSR-Net and narrow the domain gap.

More remarkably, in retrospective patient studies our spectral reconstruction quality has surpassed that with the state-of-the-art unsupervised learning method Noise2Sim. 
In our reader study, the analysis on the grading results from the combined scores from all three radiologists has demonstrated that our proposed method is better than the standard commercial reconstruction. Encouragingly, the median values of all image quality scores are on the positive side, suggesting our reconstructions are diagnostically acceptable and preferred, despite reconstructed with only halved radiation dose.
Furthermore, our method has been evaluated in terms of VGC points from the radiologists' scores against the existing method.
By each of the image quality measures our method has produced significantly better results in almost all aspects. For the total combined image quality scores, the proposed method has shown its competitive advantage against the existing method, with a mean $\text{AUC}_{VGC}$ value exceeding 0.5, although additional patient data is needed to establish statistical significance. Notably, the individual scores from all three radiologists, as well as the combined rankings, exhibit consistent trends. All favor the proposed method for a majority of the VGC points, as shown in Fig.~\ref{fig: final graph B}, but with the VGC fitting curve crossing the diagonal line at the rightmost part. This is a pattern typically associated with a method that receives higher mean scores but also exhibits greater variation in scores than the reference. This interpretation is supported by the summary statistics in Table~\ref{Table:DesSta}, where the combined ratings for the proposed half-view method show both a higher mean and a larger interquartile range compared to that of the full-view reference.
Finally, a slight to fair agreement has been obtained among radiologists, despite no formal training on their visual grading.

Regarding the radiation dose of a patient's wrist scan, we have performed comparison experiments with a GE Discovery CT750 HD scanner for standard musculoskeletal imaging. The Extremity 5x120 scanner delivers a radiation dose (volumetric CT dose index, CTDI$_{\text{vol}}$) of \(7.87mGy\) to a \(10cm\) polymethyl methacrylate (PMMA) phantom during a routine wrist scan. This radiation dose remains the same regardless of whether or not a metal implant is present in the wrist of a patient. The radiation dose of the conventional CT scanner, using wrist protocols derived by the School of Medicine and Public Health, University of Wisconsin-Madison, USA, in conjunction with GE engineers, was also measured using a \(10cm\) PMMA phantom. The radiation dose was measured at \(43.02mGy\) in CTDI$_{\text{vol}}$ using a protocol optimized for a wrist without metal implants, and \(197.3mGy\) for a protocol optimized with a metal implant present. These results suggest that the MARS scanner can produce diagnostically useful images at only a small fraction of the radiation dose required by traditional CT scanners. Moreover, our proposed method can further reduce the dose by half for improved safety without sacrificing image quality.

This study also confirms that high-quality images suitable for musculoskeletal diagnosis can be acquired in a reasonable time with a low X-ray dose at the point of care. As demonstrated in Fig.~\ref{Fig:RecResultOfPatient}(a), a clear image of a scaphoid fracture was captured with quality comparable to that of full field-of-view CT scans at only a fraction of the radiation dose. Traditional point-of-care musculoskeletal scanners typically rely on cone-beam flat panel technology, whereas the system used for this research was a photon counting scanner with an AI-based reconstruction and a helical scanning geometry. Significantly higher spatial resolution is achieved with the MARS PCDs than with the flat panel systems, along with other benefits from photon-counting technology---such as reduced metal artifacts, multi-contrast imaging, and lower radiation dose. The in-plane resolution at \(10\%\) of the modulation transfer function (MTF) peak was measured to be \(1.8 lp/mm\), while the longitudinal resolution was around \(5.0 lp/mm\)~\cite{gallego2023clinical}. These values are comparable or better than those reported for the latest Naeotom Alpha PCCT scanner from Siemens (in-plane \(1.69 lp/mm\) and axial \(3 lp/mm\) at MTF $10\%$)~\cite{thomsen2022effective}. Compared to the Siemens whole body PCCT, the MARS scanner offers unique advantages as a point-of-care solution for musculoskeletal imaging. For example, full-field PCCT scanners require substantial floor space in a radiation-shielded environment, at least two trained operators, and carry a high capital cost. With these characteristics, they are usually only installed in tertiary imaging centers, which limits patient accessibility, delays time to diagnosis, and increases cost compared to point-of-care scanners. This is especially prominent for musculoskeletal patients who are often evaluated in outpatient clinics away from the main hospital. In contrast, a MARS point-of-care scanner is more compact, requires fewer operating staff, and is well-suited for deployment in outpatient facilities where musculoskeletal assessments commonly take place.

This study is not without limitations. First, the number of patients included is relatively small, since our clinical trial is still in progress. Second, the inter-reader agreement between radiologists is low. This can be attributed to the fact that the radiologists were not trained for the inter-rater agreement regarding image quality evaluation prior to the review study. At that moment, no protocol was established in the context of PCCT, and pre-evaluation training could potentially introduce image biases. In addition, the radiologists have different degrees of familiarity with and knowledge of PCCT images. Considering the significantly lower radiation dose ($5\%$ of an equivalent traditional CT scanner protocol), they could have very different opinions on permeability to satisfactory diagnosis.
It is also worth mentioning that motion correction has been applied to five patients with noticeable movements using the method described in~\cite{li2022motion} prior to our reconstruction, which has significantly improved motion artifact assessment as shown in Fig.~\ref{fig: final graph B}(b). 

Finally, we would like to emphasize that this study aims to address the training data scarcity issue for HR volumetric PCCT reconstruction by bridging the gap between synthetic and clinical data through our proposed DIR framework. It is generally preferable to use in-domain clinical data for network training whenever possible, given they are of sufficient quality and quantity. However, such data are often scarce in practice, which sometimes necessitates the use of synthetic datasets for training. Synthetic data offer two key benefits: (1) they are relatively easy to acquire compared to patient data, and (2) they provide greater flexibility in controlling acquisition conditions, such as generating noise-free labels or producing repetitive scans under varying dose levels and motion states. On the other hand, the primary drawbacks of synthetic data are the inevitable unrealistic aspects, \emph{e.g.}, fake structures or features, simplification or omission of certain physics effects during image formation, which can introduce domain gaps and degrade network performance when applied to real clinical data. Our DIR framework bridges the domain gap through two strategies: (1) Narrowing the domain gap by training a structure-agnostic, low-level denoiser (VSR-Net); (2) Decomposing the difficult ``long jump'' into easier ``segmental walks'', with each cornerstone (intermediate result) kept quasi in-domain for the denoiser. Specifically, by designing VSR-Net as a low-level denoiser, the primary domain gap between synthetic and clinical data is reduced to differences in noise levels and image contrast modulation. To address this, by properly selecting the number of SIRT iterations when generating the structural prior, we can closely match the resulting noise magnitude to that of the synthetic data, minimizing the domain gap for VSR-Net at the first step in the DIR pipeline. In subsequent steps, we can balance the noise introduced by gradient descent updates and the noise removed by VSR-Net (controlled by the parameter $\mu$). When tuned properly, these opposing effects largely cancel each other out, maintaining a quasi in-domain input for the denoiser (note that denoisers are tolerant to cleaner images). Additionally, the image contrast modulation mismatches can be corrected through iterative feedback during DIR iterations, further ensuring the robustness of the DIR framework.

\section{Conclusion}
In conclusion, we have developed a novel deep learning method for few-view HR PCCT volumetric reconstruction in the New Zealand clinical trial at halved radiation dose and doubled imaging speed. Compared to the standard commercial reconstruction method used in the clinical trial, the proposed method produces equivalent or superior image quality at halved radiation dose. We plan to translate the proposed method for few-view image reconstruction into the PCCT system and keep improving the method as the clinical trial proceeds.

\appendix
\renewcommand{\thefigure}{A\arabic{figure}}
\setcounter{figure}{0}
A visual comparison example (an adjacent slice from that displayed in Fig.~\ref{fig:RFCAN}(a)) is displayed in Fig.~\ref{fig:SuppFigure}, as a complementary to Fig.~\ref{Fig:RecResultOfPatient}. It shows RFCAN output, final half-view reconstruction result, and full-view clinical reference. The bone edge pinpointed by the red arrow appears a little over-smoothed in the RFCAN result compared to the clinical reference. The final few-view reconstruction result improves the sharpness, which is also demonstrated in the intensity profiles. The noise texture improvement in soft tissue region is also observed, cross-validating the result observed in Figs.~\ref{Fig:RecResultOfPatient}(c) and (d). These improvements confirm the benefits of the other step from our two-step textural appearance tuning procedure besides the RFCAN processing.

\begin{figure}[h]
    \centering
    \includegraphics[width=0.98\linewidth]{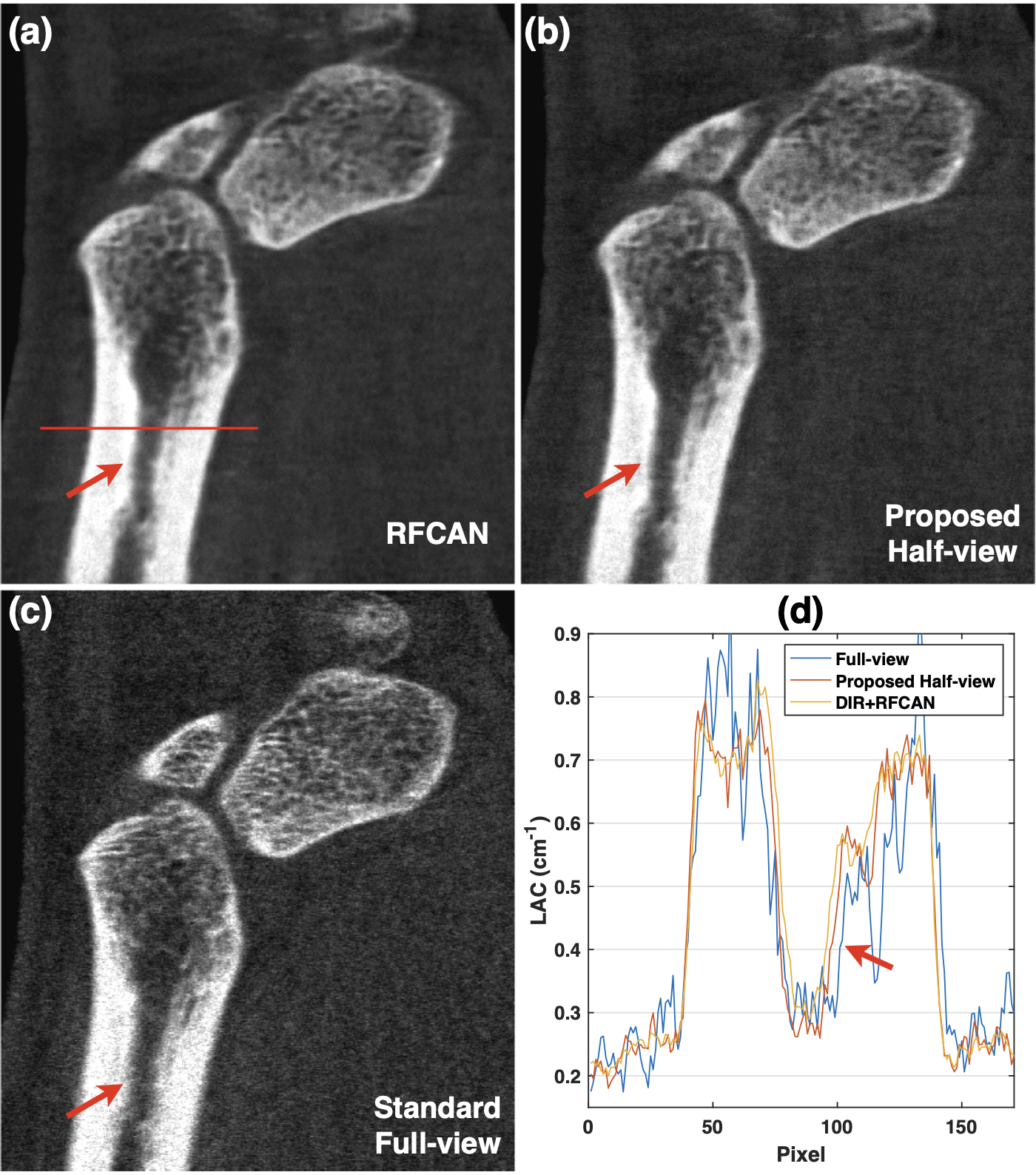}
    \vspace{-3truemm}
  \caption{Visual comparison between (a) intermediate RFCAN result, (b) final half-view reconstruction, and (c) MARS full-view reference in the channel 7-40keV, displayed in W/L:0.72/0.5 cm$^{-1}$, accompanied by (d) plots of their intensity profiles along the red line. Note the bony structure differences in the upper part of images are due to misalignment, as explained in Fig.~\ref{fig:RFCAN}(a).}
    \label{fig:SuppFigure}
    \vspace{-7truemm}
  \end{figure}

\bibliographystyle{IEEEtran}
\input{DIR_FVspectralCT.bbl}

\end{document}